\documentclass[11pt]{article}
\usepackage{amsmath}
\usepackage{amssymb,bm,epsf,epsfig,graphicx,subfigure}
\input epsf.sty
\numberwithin{equation}{section}

\usepackage{hyperref}

\usepackage{cite}
 \newcommand{\eom}		{equations of motion }
 
 \newcommand{\ls}		{\tilde\lambda}
 \newcommand{\p}		{\partial}
 \newcommand{\corr}[1]		{\left\langle\,#1\,\right\rangle}
 \newcommand{\braket}[2]	{\langle #1|#2\rangle}
 \newcommand{\bra}[1]	 	{\langle #1|}
 \newcommand{\ket}[1]		{|#1\rangle}
 \newcommand{\mth}[1]		{\mathcal{#1}}
 \newcommand{\vk}		{\ket{0}}

 \newcommand{\due}		{\tfrac{1}{2}}
 \newcommand{\mx}		{\mbox}
 \newcommand{\tin}[1]		{\mx{\tiny #1 }}
 \newcommand{\bLc}		{\frac{\cB_0}{\cL_0}}
 \newcommand{\comm}[2]		{\left[ #1,#2\right]}
 \newcommand{\Comm}[2]		{\Big[ #1,#2\Big]}
 \newcommand{\dcomm}[2]		{\Big[\!\Big[ #1,#2\Big]\!\Big]}
 \newcommand{\V}		{\mth{V}_m}
 \newcommand{\Vaux}		{\mth{V}_{\tin{aux}}^{(1-h_m,1-h_m)}}
 \newcommand{\sq}		{\sqrt{2}}
 \newcommand{\lr}[1]		{\left(#1 \right)}
 \newcommand{\z}		{\bar z}
 \newcommand{\Q}		{Q_B}

 \newcommand{\nocontentsline}[3]{}
\newcommand{\tocless}[2]{\bgroup\let\addcontentsline=\nocontentsline#1{#2}\egroup}

\usepackage{tikz}
\usetikzlibrary{calc}
\usepackage{feynmf}
\usetikzlibrary{positioning,arrows}
\usetikzlibrary{decorations.pathmorphing}
\usetikzlibrary{decorations.markings}

\tikzset{
 particle/.style={draw=black, postaction={decorate},
    decoration={markings,mark=at position .5 with {\arrow[blue]{}}}},
 invisible/.style={draw=white, postaction={decorate},
    decoration={markings,mark=at position .5 with {\arrow[blue]{}}}},
 gluon/.style={draw=blue,decorate, 
    decoration={coil,aspect=0}}
 }

\newcommand{\cB}{{\mathcal{B}}}

\newcommand{\cL}{{\mathcal{L}}}

\newcommand{\cV}{{\mathcal{V}}}

\begin{document}

\begin{center}
 {\large \bf $\,$\\
\vskip2cm
BCFT and OSFT moduli: an exact perturbative comparison}

\vskip 1.1cm

{\large Pier Vittorio Larocca\footnote{Email:
plarocca at to.infn.it} and Carlo Maccaferri \footnote{Email:
maccafer at to.infn.it}}
\vskip 1 cm
{\it Dipartimento di Fisica, Universit\'a di Torino and INFN, Sezione di Torino\\
Via Pietro Giuria 1, I-10125 Torino, Italy}\\

\end{center}

\vspace*{6.0ex}

\centerline{\bf Abstract}
\bigskip
Starting from the  pseudo-$\cB_0$ gauge solution for marginal deformations in OSFT,  we analytically compute  the  relation between the perturbative deformation parameter $\ls$ in the solution and  the BCFT marginal parameter $\lambda$, up to  fifth order,  by evaluating the Ellwood invariants.
We observe that the microscopic reason why $\ls$ and $\lambda$  are different is that the OSFT propagator renormalizes  contact term divergences differently from the contour deformation used in BCFT.  \vfill \eject

\baselineskip=15pt

\tableofcontents


\section{Introduction and conclusion}

In the recent years there has been overwhelming   evidence that the various consistent open string backgrounds ({\it i.e. }D-branes) can be described analytically as solitons  of open string field theory (OSFT) \cite{Schnabl:2005gv,Schnabl:2007az, KORZ, FKP, KO, BMT, Murata2,Erler:2012qn,Maccaferri:2014cpa, Erler:2014eqa,defects}\footnote{See \cite{Fuchs:2008cc,Schnabl:2010tb,Okawa:2012ica}  for reviews.}.

A classical \cite{Sen:2000hx} yet not fully understood problem in this correspondence is how the D-branes moduli space is described in OSFT. Given an exactly marginal boundary field $j$,  there is a corresponding family of OSFT  solutions, which can be generically found in powers of a deformation parameter $\ls$
\begin{equation}
 \Psi_{\ls} \ =\ \ls\ cj(0)\vk \ +\sum_{k=2}^{\infty}\ls^k \,\Psi_k \label{sol1},
 \end{equation}
where $\Psi_k$ are perturbative contributions obeying the recursive relation
\begin{equation}
Q\Psi_k+\sum_{n=1}^{k-1}\Psi_n\Psi_{k-n}=0.
\end{equation}

Physically we expect that the deformation parameter $\tilde\lambda$ which we used to construct the solution should be related  to the natural parameter $\lambda$ in boundary conformal field theory (BCFT),  given by the coefficient in front of the boundary interaction which deforms the original world-sheet action
\begin{equation}
S_\lambda\ =\ S_0\ +\ \lambda \int_{-\infty}^{\infty}dx\ j(x) \ .
\end{equation}

 On general grounds,  $\ls$ does not have a gauge invariant meaning, but nonetheless  it is  useful to understand how  $\lambda$ and $\tilde\lambda$ are related for a given solution, because this can shed light on the different mechanisms by which a classical  solution  changes the worldsheet boundary conditions. 

Analytic solutions for marginal deformations with nonsingular OPE ($j j \sim \mx{reg}$) have been computed to all orders in \cite{Schnabl:2007az} and \cite{KORZ}.
A different perturbative analytic solution for marginal currents with singular OPE has been constructed in \cite{FKP} and generalized in \cite{KO}\footnote{See also \cite{Karczmarek:2014wea,Longton:2015maa}.}.
An analytic solution  for any self local (hence exact \cite{Recknagel:1998ih}) marginal deformation has been constructed nonperturbatively in \cite{Maccaferri:2014cpa}. Conveniently, this solution is directly expressed in terms of the deformation parameter of the underlying BCFT, $\lambda$.
In \cite{Maccaferri:2015cha} this has been used to explicitly find the relation between the BCFT modulus $\lambda$ and the coefficient of the marginal field in the solution 
$\braket{c\p cj}{\Psi(\lambda)} $.
It has been observed  that this function of $\lambda$ starts linearly, then it has a local maximum and finally it approaches zero for large values of $\lambda$. 
Nontrivial evidence that  this behaviour may also be present in Siegel gauge\footnote{In Siegel gauge the perturbative coefficient $\ls$ \eqref{sol1} and the coefficient of the marginal field in the 
solution $\langle c\partial c j|\Psi\rangle$ coincide. This is not generically true for other perturbative solutions, see for example \cite{KO, Longton:2015maa}. This is also not true for the solution \cite{Schnabl:2007az} analysed in this paper and the relation can be computed, if needed, by the same methods of section 4. } has been given in \cite{Kudrna:2016ack} in level truncation, but it has not been possible there to establish the validity of the full \eom\ for large BCFT moduli.

In this note we would like to study this problem in  another analytic wedge-based example which is quite close to Siegel gauge.
We will analyze the observables of the solution proposed by Schnabl in  \cite{Schnabl:2007az}, in the so-called pseudo-$\cB_0$ gauge
\begin{eqnarray}
\Psi_{\ls}&=&\sum_{k=1}^{\infty} \ls ^k\ \widehat U_{k+1}\,\hat\Psi_k\ket 0,\\
\cB_0\hat\Psi_k\ket 0&=&0,
\end{eqnarray}
 for a chiral marginal current  $j(z)$ with OPE
\begin{equation}
j(z)\ j(0)\ =\ \frac{1}{z^2}\ +\ {\rm regular}.
\end{equation}
Computing the Ellwood invariants and matching them with the BCFT expected answers, gives the following relation
\begin{equation}
 \ls\ =\ls(\lambda)\ = \lambda\ -\ 3\log 2 \ \lambda^3\ +2.38996(7)\ \lambda^5 +O(\lambda^7)\ .\label{relationn}
\end{equation}
Let us comment on the found relation.

Perhaps the most interesting fact about \eqref{relationn} is the origin itself of the found coefficients of $\lambda^{2n+1}$. These coefficients are obtained by comparing the Ellwood invariants computed from the solution in powers of $\tilde \lambda$, with the coefficients of the Ishibashi states obtained from the marginally deformed boundary state expressed in powers of $\lambda$, see eqs \eqref{b0}--\eqref{b5}. Naively these two quantities reduce to the same worldsheet calculation and therefore one would expect to find perfect match between $\lambda$ and $\tilde\lambda$, which is evidently not true. This is explained as follows.
At order $\ls^k$, the encountered Ellwood invariants have the structure of OSFT  tree-level amplitudes between an on-shell closed string and $k$ on-shell  open strings given by the marginal field $cj$, with $\ls$ playing the role of the open string coupling constant. These amplitudes are naively affected by infrared divergences due to the collisions of the marginal fields at zero momentum,  which correspond to the the  propagation of the zero momentum tachyon. The  propagator $\bLc$ gives a uniquely defined prescription to renormalize these singularities, see section \ref{sec:reg}. On the other hand, in BCFT,  the  same contact term divergences are renormalized by contour deformation \cite{Recknagel:1998ih}, so that  the renormalized boundary interaction $e^{-\lambda \oint ds j(s)}$ acquires a  topological nature.  This difference in the renormalization procedure of contact term divergences is the ultimate reason why $\lambda$ and $\ls$ are different. Had the self-OPE between the currents been regular, we would have found no difference between the two quantities. 

We also observe that the growing of the coefficients in \eqref{relationn} is in agreement with the findings  from other non-perturbative approaches (although in different gauges) such as \cite{Maccaferri:2015cha} and  \cite{Kudrna:2016ack}, and it suggests that the power series in $\tilde\lambda$ may have a finite radius of convergence. It would be desirable to improve our calculation to be able to estimate the growing of the higher order coefficients and the nature of the singularity in the complex $\tilde\lambda$ space. This would be a complementary (perturbative) way of understanding the reason why (in Siegel gauge) the marginal solution breaks down at a critical value of $\tilde\lambda$. Indeed, it turns out that our computations in pseudo ${\cal B}_0$-gauge can be  related to the analogous computations in Siegel gauge, whose direct evaluation is notoriuosly very complicated. Work in this direction is in progress \cite{KLM}.  

The paper is organized as follows.
In section 2  we  review the needed material for constructing the boundary state in BCFT \cite{Recknagel:1998ih} and in OSFT  \cite{Kudrna:2012re}. Then we review the construction of the marginal solution in the pseudo-$\cB_0$ gauge \cite{Schnabl:2007az}, and we explicitly write it down up to the fifth order.
Section 3 describes  the regularization procedure implemented by the propagator $\bLc$. In section 4 we write down the coefficients of the Ishibashi states in the boundary state in terms of the deformation parameter $\lambda$ using the standard BCFT prescription by Recknagel and Schomerus \cite{Recknagel:1998ih}. Then we compute the same quantities for the OSFT solution in the pseudo-$\cB_0$ gauge.
Finally we compare the coefficients of the Ishibashi states  in OSFT and BCFT and we obtain the function $\ls=\ls(\lambda)$ up to fifth order.
An appendix  contains useful formulas for the encountered correlators.
%


\section{The boundary state and the marginal solution}

Let us consider a deformation of a BCFT by a boundary primary operator $j(x)$ of conformal weight one
\begin{equation}
 \delta S_{\tin{BCFT}}\ =\ \lambda \int j(x)\ dx \ .
\end{equation}
From the OSFT point of view the new theory can be described by a classical solution, a state in the original BCFT
\begin{equation}
 \Psi_{\ls} \ =\ \ls \Psi_1 \ +\ O(\ls^2)\ ,
\end{equation}
where
\begin{equation}
 \Psi_1 \ =\ cj(0)\ \vk \ \equiv\ \ket{cj} \ .
 \label{initial}
\end{equation}
The leading term in $\ls$ satisfies the linearized \eom
\begin{equation}
 \Q \Psi_1 \ =\ 0 \ .
\end{equation}
If $j$ is exactly marginal higher orders in $\ls$ should exist 
\begin{equation}
 \Psi_{\ls} \ =\ \sum_{k=1}^\infty \ls^k\ \Psi_k \ ,
\end{equation}
and they can be found by solving the recursive \eom
\begin{equation}
 \Q \Psi_{\ell} \ =\ \sum_{k=1}^{\ell-1} \Psi_k \Psi_{\ell-k} \ ,\label{eom-pert}
\end{equation}
with the initial condition \eqref{initial}. 

Notice that while in BCFT the perturbation is unique, the OSFT  solution is not unique because it can be changed by gauge transformations.
We can get rid of this gauge redundancy by computing observables.
In particular the information on the marginal deformation can be effectively cast in the boundary state. 

Boundary states in bosonic string theory can be written as a superposition of Ishibashi states $\ket{{\mth{V}^m}}\!\rangle$ \cite{Ishibashi:1988kg}
\begin{equation}
 \ket{B}\ =\ \sum_m n_m\ \ket{{\mth{V}^m}}\!\rangle \ \otimes \ \ket{B_{\tin{gh}}} ,
 \label{Bgh}
\end{equation}
where $\ket{B_{\tin{gh}}}$ is the universal ghost part.
When we deform a given worldsheet theory with an exactly marginal boundary deformation, the boundary state will be deformed to
\begin{equation}
 \delta S_{\tin{BCFT}} \ =\ \lambda \int j(x) \ dx\ \longrightarrow \ \ket{B(\lambda)} \ ,
\end{equation}
with
\begin{equation}
 \ket{B(\lambda)}\ =\ \left[ e^{-\lambda \oint ds\ j(s)} \right]_{\tin{R}} \ket{B_0} \ ,
\end{equation}
where $\left[\ldots\right]_{\tin{R}}$ means that a regularization is needed (and it will be reviewed later on), and $\ket{B_0}$ is the boundary state of the starting BCFT.

On the other hand, given an OSFT solution $\Psi_{\ls}$, the boundary state will depend on $\ls$ 
\begin{equation}
 \Psi_{\ls} \longrightarrow\ \ket{B_{\Psi}(\ls)} \ .
\end{equation}
The two boundary states should be the same by the Ellwood conjecture \cite{Ellwood:2008jh} and this induces a functional relation
\begin{equation}
 \ls \ =\ \ls (\lambda) \ .
\end{equation}
To obtain this relation we can compare the coefficients of the Ishibashi states.
From \eqref{Bgh} it follows that\footnote{From now on we will only consider the matter part of boundary states.}
\begin{equation}
 n_m^{\tin{BCFT}}(\lambda)\ =\ \braket{\V}{B(\lambda)}\ =\ \bra{\V } \left[ e^{-\lambda \oint ds\ j(s)} \right]_{\tin{R}} \ket{B_0} =\left\langle   \left[ e^{-\lambda \oint ds\ j(s)} \right]_{\tin{R}} \V(0,0)\right\rangle_{\tin{Disk}}\ ,
 \label{paragoneIshibashi}
\end{equation}
where $\bra{\V}$ is the BPZ conjugate of the matter Virasoro primary $\ket{{\cal V}^n}$ so that 
$\braket{\V}{{\cal V}^n}=\braket{\V}{{\cal V}^n}\!\rangle=\delta_m^n$
where we used the fact that Ishibashi states have the generic form
\begin{equation}
\ket{{\cal V}^n}\!\rangle=\ket{{\cal V}^n}+\textrm{Virasoro descendants}.
\end{equation}
The series expansion of the exponential in (\ref{paragoneIshibashi}) gives rise to contact divergences and one needs to renormalize them properly. In the next section we will review the standard procedure of \cite{Recknagel:1998ih}.

The way to compute the $n_m$'s from OSFT was given in \cite{Kudrna:2012re} by appropriately generalising the Ellwood invariant 
\begin{equation}
 n_m^{\tin{SFT}}(\ls)\ =\ \braket{\V}{B_{\Psi}(\ls)}\ =\ 2 \pi i\ \bra{\mth{I}} \ V_m^{(0,0)}(i,-i)\ \ket{\Psi_{\ls}-\Psi_{\tin{TV}}} \ ,
 \label{EllwoodSFT}
\end{equation}
where $\Psi_{\tin{TV}}$ is a tachyon vacuum solution and $V_m^{(0,0)}(i,-i)$ is a weight $(0,0)$ bulk field of the form
\begin{equation}
 V_m^{(0,0)} \ \equiv\ c\bar c\ V_m^{(1,1)} \ =\ c\bar c\ \cV_m^{(h_m,h_m)} \ \otimes\ \cV_{\tin{aux}}^{(1-h_m,1-h_m)} .
 \label{VauxVmat}
\end{equation}
As explained in detail in \cite{Kudrna:2012re}, the auxiliary bulk field $\cV_{\tin{aux}}^{(1-h_m,1-h_m)} $ lives in an auxiliary $\mx{BCFT}_{\tin{aux}}$of $c=0$ and has unit one-point function on the disk
\begin{equation}
 \corr{\cV^{(1-h_m,1-h_m)}_{\tin{aux}}(0,0)}_{\tin{Disk}}\ =\ 1 \ .
 \label{Vaux}
\end{equation}
In a similar way the open string fields entering in \eqref{EllwoodSFT} are lifted to the  extended BCFT
\begin{equation}
 \mx{BCFT}_0^{\tin{new}}\ =\ \mx{BCFT}_0\ \otimes\ \mx{BCFT}_{\tin{aux}}\ .
\end{equation}
For the solution we will be dealing with this lifting procedure is trivial and amounts to the substitution $L_n\to L_n+L_n^{(\rm aux)}$ in the equations that will follow. For this reason we will not distinguish between normal and lifted string fields in the sequel.

As far as the solution itself is concerned, we search for it in the convenient pseudo-$\cB_0$ gauge \cite{Schnabl:2007az}, making the following ansatz
    \begin{equation}
     \Psi_{\ls} = \sum_{r=1}^\infty (\ls)^r\ U_{r+1}^* U_{r+1} \hat\Psi_{r} \vk \ ,
    \end{equation}
    with the gauge condition
    \begin{equation}
     \cB_0 \hat\Psi_r \ \vk \ =\ 0 \ ,
    \end{equation}
    where $\cB_0$ is the zero mode of the $b$ ghost in the sliver frame, obtained from the UHP by the conformal transformation $z=\frac2\pi\arctan w$
    \begin{equation}
     \cB_0\ =\ \oint \frac{dz}{2\pi i}\ z\ b(z) \ ,
    \end{equation}
and the operators $U_r$ are the common exponentials of total Virasoro operators creating the wedge states  \cite{Schnabl:2002gg} in the well known way
\begin{equation}
 \ket{r}\ =\ U_r^* U_r \vk\ =\ U_r^* \vk\ =\ \underbrace{\vk * \ldots * \vk }_{r-1} \ .
\end{equation}
 Solving order by order in $\ls$ \eqref{eom-pert} we find
\begin{itemize}
 \item[$O(\lambda^2)$:] $ Q\Psi_2 = -(cj)^2$. \\
 The rhs is explicitly given by
        \begin{equation}
     (cj)^2\ \equiv\ cj(0)\vk * cj(0)\vk \ =\ U_3^* U_3\ cj\lr{\due}cj\lr{-\due} \vk \ ,
    \end{equation}
    where the $cj$ insertions are written in the sliver frame. 
    The solution is therefore
    \begin{equation}
     \Psi_2 \ =\  U_3^* U_3\ \hat\Psi_2 \vk \ =\ -\ U_3^* U_3\ \bLc\ cj\lr{\due} cj\lr{-\due}\vk \ ,
     \label{psi2}
    \end{equation}
    where $\cL_0$ is the zero mode of the energy-momentum tensor in the sliver frame,
    \begin{equation}
     \cL_0\ =\ \oint \frac{dz}{2\pi i}\ z\ T(z) \ .
    \end{equation}
    Note that inverting $\Q$ using $\cB_0/\cL_0$ is only meaningful if the OPE of $cj$ with itself does not produce weight zero terms, otherwise we would find a vanishing eigenvalue of $\cL_0$.  As is well known this is the first nontrivial condition for $j$ to generate an exactly marginal deformation.\\
 \item[$O(\lambda^3)$:] $\Q\Psi_3 + \comm{cj}{\Psi_2} =0$. \\
    At the third order the solution $\Psi_3$ is written in terms of $\Psi_2$.
    We write  the state $\comm{cj}{\Psi_2}$  as
    \begin{equation}
     \begin{split}
      \comm{cj}{\Psi_2}\ &=\ \Comm{ U_2^* U_2\ cj(0)\vk }{\ U_3^* U_3\ \hat\Psi_2\vk }\ \\
      &=\ U_4^* U_4\ \dcomm{cj(0)}{\hat\Psi_2}_{\tin{(2,3)}} \vk\ \ ,
     \end{split}
  \end{equation}
    where in the second step we explicitly write the width of the wedge states using the $U_r$ operators. The inside insertions have to be placed according to  \eqref{starSchnabl}: to lighten a bit the notation we have defined the graded commutator-like symbol $[\![\ldots]\!]$ as
    \begin{equation}
     \dcomm{\psi(x)}{\phi(y)}_{\tin{(r,s)}}\ \equiv\ \psi\lr{x+\tfrac{s-1}{2}} \phi\lr{y-\tfrac{r-1}{2}}\ -\ (-1)^{|\psi||\phi|}\ \phi\lr{y+\tfrac{r-1}{2}} \psi\lr{x-\tfrac{s-1}{2}} \ ,
    \end{equation}
    coming from
    \begin{equation}
     \comm{ U_r^* U_r\ \psi(x)\vk }{\ U_s^* U_s\ \phi(y)\vk }\ =\ U_{r+s-1}^* U_{r+s-1}\ \dcomm{\psi(x)}{\phi(y)}_{\tin{(r,s)}}\ \vk \ .
    \end{equation}
    Finally we can write the solution at the third order as
    \begin{equation}
     \Psi_3\ =\ U_4^* U_4\ \hat\Psi_3 \vk\ =\ -\ U_4^* U_4\ \bLc\ \dcomm{cj}{\hat\Psi_2}_{\tin{(2,3)}} \vk\ \ .
     \label{psi3}
    \end{equation}

 \item[$O(\lambda^4)$:] $\Q\Psi_4 + \comm{cj}{\Psi_3} + \Psi_2^2 =0$. \\
    Again,
    \begin{equation}
     \Psi_4 \ =\   U_5^* U_5\ \hat\Psi_4 \vk\ =\ -\ U_5^* U_5\ \bLc\ \left( \frac{1}{2}\ \dcomm{\hat\Psi_2}{\hat\Psi_2}_{\tin{(3,3)}}  \ +\ \dcomm{cj}{\hat\Psi_3}_{\tin{(2,4)}}  \right) \vk\ .
     \label{psi4}
    \end{equation}
  \item[$O(\lambda^5)$:] $\Q\Psi_5 + \comm{cj}{\Psi_4} + \comm{\Psi_2}{\Psi_3} =0$. \\
    We find the fifth order as
    \begin{equation}
     \begin{split}
      \Psi_5\ &=\ U_6^* U_6\ \hat\Psi_5 \vk\ \\
      &=\ -\ U_6^* U_6\ \bLc\ \Bigg( \dcomm{cj}{\hat\Psi_4}_{\tin{(2,5)}} \ +\ \dcomm{\hat\Psi_2}{\hat\Psi_3}_{\tin{(3,4)}} \Bigg) \vk\ \ .
     \end{split}
    \end{equation}
\end{itemize}
\bigskip\bigskip
This procedure can be continued to higher order\footnote{In \cite{Schnabl:2007az} an all-order expression is written down, by explicitly acting with the $\cB_0$ ghosts, which coincides with our expressions, up to the issue of  renormalizing contact term divergences. We keep the propagators $\cB_0/\cL_0$ explicit, because they are the origin of the contact-term renormalization we will describe next.}. Although higher orders can be easily written down,  their Ellwood invariants become more and more complicated because they involve a large number of multiple integrals which by themselves need to be properly renormalized, as we will see in the next section.\\


\section{Contact term divergences and the propagator}
\label{sec:reg}
The computation of the Ellwood invariants for the solution we have just presented involves in general contact divergences due to the definition of the propagator $\cB_0/\cL_0$. As usual, we start by defining the inverse of $\cL_0$  via the Schwinger representation
\begin{equation}
 \frac1\cL_0\ =\ \int_0^1 \frac{ds}{s}\ \ s^{\cL_0} \ ,\label{Schwinger}
\end{equation}
which is well defined for eigenvalues of $\cL_0$ with a strictly positive real part.
The operator $s^{\cL_0}$ is the generator of dilatations $z\mapsto s z$ in the sliver frame. Its action on a primary field  with conformal weight $h$ in the sliver frame is
\begin{equation}
s^{\cL_0} \phi(z)= s^h\ \phi\lr{s z} \ .
\end{equation}
The integral representation (\ref{Schwinger}) is only valid for fields with a positive scaling dimension $h>0$, 
if we apply the above integral representation  to a state $\ket{\varphi_{-|h|}}$ with negative weight $-|h|$
\begin{equation}
 \frac{1}{\cL_0}\ \ket{\varphi_{-|h|}}\ =\ \int_0^1 \frac{ds}{s}\ s^{\cL_0}\ \ket{\varphi_{-|h|}}\ =\ \ket{\varphi_{-|h|}} \int_0^1 \frac{ds}{s^{|h|+1}},\quad\quad  \textrm{(incorrect)}\nonumber
\end{equation}
 we find a divergence, as $s$ approaches zero. But this is just the reflection that  the integral representation \eqref{Schwinger} has been used outside its domain of validity. This can be easily remedied in the following way
\begin{equation}
 \frac{1}{\cL_0}\ =\ \left. \frac{1}{\cL_0+\epsilon} \ \right|_{\epsilon = 0} =\ \left. \int_0^1 \frac{ds}{s^{1-\epsilon}}\ s^{\cL_0}  \ \right|_{\epsilon = 0} \ .
 \label{propagator}
\end{equation}
This prescription amounts to computing the Schwinger integral in its region of convergence by assuming ${\rm Re}(\epsilon)> |h|$, and then analytically continuing to $\epsilon=0$ \footnote{An equivalent prescription is the Hadamard regularization, we thank M. Frau for  discussions on this. See also \cite{Schnabl:2007az, Berkovits:2003ny}.}

\begin{eqnarray}
 \frac{1}{\cL_0}\ \ket{\varphi_{-|h|}}\ &=&\ \left.  \int_0^1 \frac{ds}{s^{1-\epsilon}}\ s^{\cL_0}\ \ket{\varphi_{-|h|}}\right|_{\epsilon = 0}
=\left.\ket{\varphi_{-|h|}} \int_0^1 ds\ s^{\epsilon - |h| - 1}\ \right|_{\epsilon = 0}\nonumber\\
&=&\ \left[ \ket{\varphi_{-|h|}}\left. \frac{s^{\epsilon - |h|}}{\epsilon - |h|} \right|_{0}^{1}\ \right]_{\epsilon= 0} =\left.\frac1{\epsilon-|h|}\ \ket{\varphi_{-|h|}}\right|_{\epsilon=0}=-\frac1{|h|}\ \ket{\varphi_{-|h|}}\end{eqnarray}
This analytic continuation allows to define $\cL_0^{-1}$ on every state  we encounter during our computations{ \it except} on weight zero states which remain as an  obstruction, as it should be.
\footnote{Notice that one could in principle define $\cL_0^{-1}$ on negative weight states as
\begin{equation}
\frac1{\cL_0}\ket{\varphi_{-|h|}}=\frac{-1}{-\cL_0}\ket{\varphi_{-|h|}}=-\int_0^1\frac{ds}{s}s^{-\cL_0}\ket{\varphi_{-|h|}}=-\frac1{|h|}\ \ket{\varphi_{-|h|}},
\end{equation}
however this integral representation does not work for positive weight states. Since the star product generates both positive and negative weight states at the same time, we need a representation of $\cL_0^{-1}$ that works on the whole set of fields (except, of course, the weight zero fields).}
Pragmatically, this procedure is equivalent to add and remove the tachyon contribution from the OPE, for example
\begin{equation}
 \bLc\ \Big[ cj(x)\ cj(-x)\Big]\to \ \bLc\ \Big[ cj(x)\ cj(-x)\ +\ \frac{1}{2x}\ c\p c (0)\Big]\ - \bLc\ \ \frac{1}{2x}\ c\p c (0) ,
\end{equation}
and to define  $1/\cL_0$ on the tachyon as
\begin{equation}
 \frac{1}{\cL_0}\ \ket{c \p c }\ =\ -\ \ket{c\p c} \ .
\end{equation}


\section{Comparing $\lambda$ and $\ls$}

In this section we perturbatively compute the coefficients of the series expansion of  $\ls = \ls(\lambda)$
\begin{equation}
 \tilde\lambda\ =\ \ls(\lambda)\ =\ \sum_{k=0}^\infty b_k\ \lambda^k \ ,
 \label{comparing}
\end{equation}
up to fifth order. On general grounds we expect that $b_0=0$ and $b_1=1$, and this will be verified in the next subsections.
The $b_k$'s are computed by equating  the coefficients of the Ishibashi states in the boundary state in BCFT and OSFT  \cite{Kudrna:2012um,  Kudrna:2012re}
\begin{equation}
 n_m^{\tin{BCFT}}(\lambda)\ =\ n_m^{\tin{SFT}}(\ls) \ .
 \label{equating}
\end{equation}
In both cases one can expand the above coefficients in power series of the corresponding deformation parameter
\begin{equation}
 n_m^{\tin{BCFT}}(\lambda)\ \equiv \ \sum_{k=0}^\infty B_{k,m}^{\tin{BCFT}} \lambda^k  \ ,\hspace{3cm}  n_m^{\tin{SFT}}(\ls)\ \ \equiv \ \sum_{k=0}^\infty B_{k,m}^{\tin{SFT}} \tilde\lambda^k \ .
 \label{ishibashi}
\end{equation}
The $B_{k,m}^{\tin{BCFT}}$ coefficients can be found expanding the exponential in \eqref{paragoneIshibashi}
\begin{equation}
 B_{k,m}^{\tin{BCFT}} \ \equiv \ \frac{(-1)^k}{k!}\ 2^{2h_m} \int_{-\infty}^\infty ds_1 \dots \int_{-\infty}^\infty ds_k\ \corr{\V(i,-i)\ j(s_1) \dots j(s_k)}_{\tin{UHP}} \ .
 \label{renint}
\end{equation}
where the conformal factor $ 2^{2h_m}$ comes from the transformation of $\cV_m$ under the map from the disk to the UHP.\\
These integrals need a renormalization, discussed by Recknagel and Schomerus \cite{Recknagel:1998ih}: thanks to the self locality property of the current $j$, one can modify the path of each integral to be parallel to the real axis but with a positive immaginary part $\epsilon$, with $0 < \epsilon <\!< 1$,
\begin{equation}
 \int_{-\infty}^{\infty}\ dx_{k-1} \ \longrightarrow \ \int_{-\infty +ik\epsilon}^{\infty +ik\epsilon}\ dx_{k-1} \ ,
\end{equation}
In such a way all the contact divergences between the currents are avoided and only the contraction of the currents with the closed string will give contribution. Thanks to this renormalization the loop operator $\left[e^{-\lambda\oint j(s)ds}\right]_{\tin R}$ becomes a topological defect.

For the sake of simplicity we consider an exactly marginal deformation produced by the operator
\begin{equation}
 j(z)\ =\ i \sq\ \p X(z) \ ,
 \label{jdx}
\end{equation}
on an initial Neumann boundary condition of a free boson compactified at the self-dual radius $R=1$($\alpha'=1$)\footnote{Since we are considering a compactified theory at the self-dual radius, there are other marginal operators in the enlarged $SU(2)$ chiral algebra. Our choice \eqref{jdx} is equivalent to the chiral marginal operators $i\sqrt{2}\sin \lr{2X(z)}$ and $i\sqrt{2}\cos \lr{2X(z)}$, which have been studied in a similar context in \cite{Kudrna:2016ack, Kudrna:2012um} }. This deformation switches on a Wilson line in the compactified direction which can be detected by a closed string vertex operator carrying winding charge
\begin{equation}
 \V(z,\z) \ =\ e^{i m \tilde X(z,\z)} \ ,
\end{equation}
where $m$ is the winding number (which specifies the closed string index) and $\tilde X(z,\z)=X(z) - \bar X(\z)$ the T-dual field of $X(z,\z)$\footnote{If $R$ is not self dual, our computation goes on unaffected by replacing the self-dual winding mode $m$ with the winding mode at generic radius, $mR$. } . This closed string state has conformal weight $(\frac{m^2}{4},\frac{m^2}{4})$. \\
Performing the renormalized integral \eqref{renint}, one obtain with this choice of the current and closed string state (see appendix \ref{app} for conventions and basic correlators)
\begin{equation}
 B_{k,m}^{\tin{BCFT}}\ =\ \frac{\left(-\ im\sqrt{2}\ \pi \right)^k}{k!} \ ,
 \label{BCFTcoeff}
\end{equation}
which can be easily resummed to
\begin{equation}
 n_m^{\tin{BCFT}}(\lambda)\ =\ e^{-im\sqrt{2}\ \pi \lambda}\ .
\end{equation}

In the OSFT framework, the analytic computation of coefficients of the Ishibashi state involves the Ellwood invariants and we compute them order by order in $\ls$, starting from \eqref{EllwoodSFT} which gives 
\begin{equation}
 \begin{split}
  &B_{0,m}^{\tin{SFT}} \ =\ -\ 2 \pi i\ \bra{\mth{I}} \ V_m^{(0,0)}(i,-i)\ \ket{\Psi_{\tin{TV}}}\ , \\
  &B_{1,m}^{\tin{SFT}} \ =\ 2 \pi i\ \bra{\mth{I}} \ V_m^{(0,0)}(i,-i)\ \ket{cj}\ , \\
  &B_{k,m}^{\tin{SFT}} \ =\ \ 2 \pi i\ \bra{\mth{I}} \ V_m^{(0,0)}(i,-i)\ \ket{\Psi_{k}} \ ,\qquad\qquad  k\geq 2 \ .
 \end{split}
 \label{amplIshi}
\end{equation}
The relation between $\lambda$ and $\ls$ must be universal, in the sense that it cannot depend on the particular choice of the closed string. In our specific computation we will see that this is the case by verifying that the relation is independent of the winding charge $m$.
$ $\\

Now rewriting \eqref{equating} using \eqref{ishibashi} and \eqref{comparing} gives explicitly 
\begin{itemize}
 \item[$O(\lambda^0)$] \hspace{2cm} 
  \begin{equation}
  b_0 \ \propto\ B_{0,m}^{\tin{SFT}}  -B_{0,m}^{\tin{BCFT}} \ (=\ 0) \ ,
  \label{b0}
 \end{equation} 
\end{itemize}

\begin{itemize}
 \item[$O(\lambda^1)$] \hspace{2cm}
 \begin{equation}
  b_1\ =\ \dfrac{B_{1,m}^{\tin{BCFT}}}{B_{1,m}^{\tin{SFT}}} \ ,
  \label{b1}
 \end{equation} 
\end{itemize}

\begin{itemize}
 \item[$O(\lambda^2)$] \hspace{2cm}
 \begin{equation}
  b_2\ =\ \dfrac{ \lr{B_{1,m}^{\tin{SFT}}}^2 B_{2,m}^{\tin{BCFT}}\ -\ \lr{B_{1,m}^{\tin{BCFT}}}^2 B_{2,m}^{\tin{SFT}} }{ \lr{B_{1,m}^{\tin{SFT}}}^3 } \ ,
  \label{b2}
 \end{equation}
\end{itemize}

\begin{itemize}
 \item[$O(\lambda^3)$] \hspace{2cm}
 \begin{equation}
  \begin{split}
   b_3\ &=\ \dfrac{ \lr{B_{1,m}^{\tin{SFT}}}^3 B_{3,m}^{\tin{BCFT}}\ -\ \lr{B_{1,m}^{\tin{BCFT}}}^3 B_{3,m}^{\tin{SFT}} }{ \lr{B_{1,m}^{\tin{SFT}}}^4 }\ +\ \\
   &+\ 2\ \dfrac{ B_{1,m}^{\tin{BCFT}} B_{2,m}^{\tin{SFT}} }{ \lr{B_{1,m}^{\tin{SFT}}}^4 }\ \lr{ \lr{B_{1,m}^{\tin{BCFT}}}^2 B_{2,m}^{\tin{BCFT}}\ -\ \lr{B_{1,m}^{\tin{BCFT}}}^2 B_{2,m}^{\tin{SFT}} } \ ,
  \end{split}
  \label{b3}
 \end{equation}
\end{itemize}

\begin{itemize}
 \item[$O(\lambda^4)$] \hspace{2cm}
 \begin{equation}
  \begin{split}
   b_4 \ &=\ \frac{\lr{B_{1,m}^{\tin{BCFT}}}^4 \lr{ 5 B_{2,m}^{\tin{SFT}} B_{3,m}^{\tin{SFT}} -\lr{B_{1,m}^{\tin{SFT}}} B_{1,m}^{\tin{SFT}} } }{\lr{B_{1,m}^{\tin{SFT}}}^6} \ -\ 3 \frac{\lr{B_{1,m}^{\tin{BCFT}}}^2  B_{1,m}^{\tin{BCFT}} B_{3,m}^{\tin{SFT}} }{ \lr{B_{1,m}^{\tin{SFT}}}^4 } \  \\
   &+ \frac{ 6 \lr{B_{1,m}^{\tin{BCFT}}}^2 B_{2,m}^{\tin{BCFT}} \lr{B_{2,m}^{\tin{SFT}}}^2}{\lr{B_{1,m}^{\tin{SFT}}}^5} \ +\ \frac{ \lr{B_{1,m}^{\tin{SFT}}}^6 B_{4,m}^{\tin{BCFT}} -5 \lr{B_{1,m}^{\tin{BCFT}}}^4 \lr{B_{2,m}^{\tin{SFT}}}^3 }{\lr{B_{1,m}^{\tin{SFT}}}^7} \  \\
   &-\ \frac{ B_{2,m}^{\tin{SFT}} \left(2 B_{1,m}^{\tin{BCFT}} B_{3,m}^{\tin{BCFT}} +\lr{B_{2,m}^{\tin{BCFT}}}^2\right)}{\lr{B_{1,m}^{\tin{SFT}}}^3} \ ,
  \end{split}
  \label{b4}
 \end{equation}
\end{itemize}

\begin{itemize}
 \item[$O(\lambda^5)$] \hspace{2cm}
 \begin{equation}
  \begin{split}
   b_5\ &=\ \frac{14 \lr{B_{1,m}^{\tin{BCFT}}}^5 \lr{B_{2,m}^{\tin{SFT}}}^4}{\lr{B_{1,m}^{\tin{SFT}}}^9}\ -\ \frac{21 \left(\lr{B_{1,m}^{\tin{BCFT}}}^5 \lr{B_{2,m}^{\tin{SFT}}}^2 B_{3,m}^{\tin{SFT}}\right)}{\lr{B_{1,m}^{\tin{SFT}}}^8} \  \\
   &+\ \frac{6 \lr{B_{1,m}^{\tin{BCFT}}}^5 B_{2,m}^{\tin{SFT}} B_{4,m}^{\tin{SFT}} + 3 \lr{B_{1,m}^{\tin{BCFT}}}^5 \lr{B_{3,m}^{\tin{SFT}}}^2 - 20 \lr{B_{1,m}^{\tin{BCFT}}}^3 B_{2,m}^{\tin{BCFT}} \lr{B_{2,m}^{\tin{SFT}}}^3}{\lr{B_{1,m}^{\tin{SFT}}}^7}  \\
   &+\ \frac{20 \lr{B_{1,m}^{\tin{BCFT}}}^3 B_{2,m}^{\tin{BCFT}} B_{2,m}^{\tin{SFT}} B_{3,m}^{\tin{SFT}}-\lr{B_{1,m}^{\tin{BCFT}}}^5 B_{5,m}^{\tin{SFT}}}{\lr{B_{1,m}^{\tin{SFT}}}^6}  \\
   &+\ \frac{-4 \lr{B_{1,m}^{\tin{BCFT}}}^3 B_{2,m}^{\tin{BCFT}} B_{4,m}^{\tin{SFT}}+6 \lr{B_{1,m}^{\tin{BCFT}}}^2 \lr{B_{2,m}^{\tin{SFT}}}^2 B_{3,m}^{\tin{BCFT}}+6 B_{1,m}^{\tin{BCFT}} \lr{B_{2,m}^{\tin{SFT}}}^2 \lr{B_{2,m}^{\tin{SFT}}}^2}{\lr{B_{1,m}^{\tin{SFT}}}^5} \  \\
   &+\ \frac{-3 \lr{B_{1,m}^{\tin{BCFT}}}^2 B_{3,m}^{\tin{BCFT}} B_{3,m}^{\tin{SFT}}-3 B_{1,m}^{\tin{BCFT}} \lr{B_{2,m}^{\tin{SFT}}}^2 B_{3,m}^{\tin{SFT}}}{\lr{B_{1,m}^{\tin{SFT}}}^4} \  \\
   &+\ \frac{-2 B_{1,m}^{\tin{BCFT}} B_{2,m}^{\tin{SFT}} B_{4,m}^{\tin{BCFT}}-2 B_{2,m}^{\tin{BCFT}} B_{2,m}^{\tin{SFT}} B_{3,m}^{\tin{BCFT}}}{\lr{B_{1,m}^{\tin{SFT}}}^3}+\frac{B_5^{\tin{BCFT}}}{B_{1,m}^{\tin{SFT}}} \ .
  \end{split}
  \label{b5}
 \end{equation}
\end{itemize}
$ $\\


\tocless\subsection{Zeroth order}
As a starting consistency check,
the zeroth order of the expansion of the coefficients of the Ishibashi states in OSFT is
\begin{equation}
 B_{0,m}^{\tin{SFT}} \ =\ -\ 2\pi i\ \bra{ \mth{I}} V_m^{(0,0)}(i,-i) \ket{ \Psi_{\tin{TV}}} \ ,
\end{equation}
where as explained in \cite{Kudrna:2012re} the tachyon vacuum contribution can be replaced with $\Psi_{\tin{TV}} \rightarrow \frac{2}{\pi} c(0) \vk$.
The amplitude then becomes\footnote{From now on we will write $V_m$ instead of $V_m^{(0,0)}$ to denote the lifted closed string state associated to the spinless matter  primary $\V$.}
\begin{equation}
 B_{0,m}^{\tin{SFT}} \ =\ -\ 4 i\ \bra{ \mth{I}} V_m(i,-i) c(0) \vk \ = \ -\ 4 i\ \corr{  V_m(i,-i)\ f_\mth{I} \circ c(0) }_{\tin{UHP}} \ ,
\end{equation}
where we used the conformal map defining the identity string field $f_\mth{I} (z) = \frac{2z}{1-z^2}$.
Then
\begin{equation}
 \begin{split}
  B_{0,m}^{\tin{SFT}} \ &=\ -\ 2 i\ \corr{c(i)c(-i)c(0) }_{\tin{UHP}} \corr{\V(i,-i)}_{\tin{UHP}} \corr{ \mth{V}_{\tin{aux}}^{(1-h_m,1-h_m)}(i,-i)}_{\tin{UHP}} \ \\
  &=\ -\ 2 i\ (2i)\ 2^{-2h_m} \ 2^{2(h_m-1)} \ =\ 1 \ .
 \end{split}
\end{equation}\\
Consistently we find
\begin{equation}
 B_{0,m}^{\tin{SFT}} = B_{0,m}^{\tin{BCFT}} =1\ ,
\end{equation}
which confirms that
\begin{equation}
 b_0\ =\ 0 \ .
\end{equation}

\tocless\subsection{First order}
As an extra starting check, let us look at the first order, where we have to compute
\begin{equation}
  B_{1,m}^{\tin{SFT}} \ = \ 2\pi i\ \bra{ \mth{I}} V_m(i,-i) \ket{cj}\ =\ 2\pi i\ \corr{ V_m(i\infty,-i\infty)\ cj(0) }_{C_1}\ ,
\end{equation}
where in the last step we wrote the correlator on a cylinder of width one $C_1$, without any conformal factor because the conformal weight of all the insertions is zero.\\
Acting with the map
\begin{equation}
 z \rightarrow\ e^{2\pi i z} \ ,
\end{equation}
this two point function on the cylinder becomes the two point function on the disk $D$,
\begin{equation}
 \begin{split}
  B_{1,m}^{\tin{SFT}} \ =\ & 2\pi i\ \corr{ V_m(0,0)\ cj(1) }_{D}\ \\
  &=\ 2\pi i\ \corr{c \bar c(0)c(1)}_{D} \corr{ \V(0,0)\ j(1) }_{D} \corr{\mth{V}_{\tin{aux}}^{(1-h_m,1-h_m)}(0,0)}_{D}\ \\
  &=\ 2\pi i\ \corr{ \V(0,0)\ j(1) }_{D}\ =\ -\ i \pi m \sqrt{2} \ ,
 \end{split}
\end{equation}
which equals the amplitude computed from BCFT \eqref{BCFTcoeff},
\begin{equation}
 B_{1,m}^{\tin{BCFT}}\ =\ B_{1,m}^{\tin{SFT}} \ ,
\end{equation}
and so the corresponding coefficient in the $\ls / \lambda$ relation is
\begin{equation}
 b_1\ =\ 1 \ .
\end{equation}

\tocless\subsection{Second order}

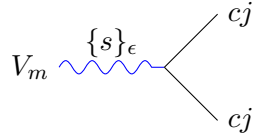
\begin{figure}[h]
 \centering
 \begin{tikzpicture}[node distance=.5cm and .5cm]
  \coordinate[label=left:$V_m$] (e1);
  \coordinate[right=1.4cm of e1] (aux1);
  \coordinate[below right=1.cm of aux1,label=right:$cj$] (e2);
  \coordinate[above right=1.cm of aux1,label=right:$cj$] (e3);
  
  \coordinate[right=.7cm of e1,label=above:$\{s\}_\epsilon$];

  \draw[gluon] (e1) -- (aux1);
  \draw[particle] (aux1) -- (e2);
  \draw[particle] (aux1) -- (e3);
 \end{tikzpicture}
 \caption{Diagram related to $\corr{V_m\hat\Psi_2 }$. The  $s$ variable is the Schwinger parameter (taking value in the interval $[0,1]$) of the propagator and $\epsilon$ is the corresponding regulator.  \label{fig:corr1}}
\end{figure}

At second order the Ellwood invariant contains one $\cB_0/\cL_0$ propagator inside $\Psi_2$,
\begin{equation}
 \begin{split}
  B_{2,m}^{\tin{SFT}} \ &=\ 2\pi i\ \bra{ \mth{I}} V_m(i,-i) \ket{\Psi_2} \ \\
  &=\ 2\pi i\ \corr{V_m(i\infty,-i\infty)\ \hat\Psi_2}_{C_2} \ \\
  &=\ - \ 2\pi i \corr{V_m(i\infty,-i\infty)\ \left( \bLc \ cj\lr{\due}\ cj\lr{-\due} \right) }_{C_2} \ .
 \end{split}
\end{equation}
This amplitude is depicted in figure \ref{fig:corr1}.

The action of the propagator on the double insertion of $cj$ follows the regularization \eqref{propagator}, so this state can be written as
\begin{equation}
 \begin{split}
  \Psi_2 \ &=\ -\ U_3^* U_3\ \bLc cj\lr{\due} cj\lr{-\due} \ \vk \ \\
  &=\ -\ U_3^* U_3 \int_0^1 \frac{ds}{s^{1-\epsilon}}\ \cB_0\ s^{\cL_0}\  cj\lr{\due} cj\lr{-\due} \ \vk\ \Bigg|_{\epsilon=0}\ \\
  &= -\ U_3^* U_3 \int_0^1 \frac{ds}{s^{1-\epsilon}}\ \cB_0\ cj\lr{\tfrac{s}{2}} cj\lr{-\tfrac{s}{2}} \ \vk \ \Bigg|_{\epsilon=0}\ ,
 \end{split}
\end{equation}
The $\cB_0$ ghost acts on $c(w)$ as the contour integral
\begin{equation}
 [\cB_0 ,c(w)]\ =\ \oint_{w} \frac{d z}{2\pi i}\ z\ b(z) c(w)=w,
\end{equation}
so that we have
\begin{equation}
 \Big[ \cB_0 , c\lr{\tfrac{s}{2}} c\lr{-\tfrac{s}{2}} \Big]\ =\ \frac{s}{2} \ \Big[ c\lr{-\tfrac{s}{2}}\ +\ c\lr{\tfrac{s}{2}} \Big] \ .
\end{equation}
Therefore (renaming $s/2\to s$)$\Psi_2$ simplifies
\begin{equation}
 \Psi_2\ =\ -\ U_3^* U_3 \int_0^{\due} ds\ s^\epsilon\ \Big[ j\lr{s} cj\lr{-s} + cj\lr{s} j\lr{-s} \Big] \ \vk \Bigg|_{\epsilon=0} \ ,
\end{equation}
and the amplitude becomes
\begin{equation}
 B_{2,m}^{\tin{SFT}} \ =\ -\ 4\pi i\ \int_0^{\frac{1}{2}} ds\ s^\epsilon\ \corr{ V_m(i\infty)\ cj(s)\ j(-s) }_{C_2} \ \ \Bigg|_{\epsilon=0} \ ,
\end{equation}
where we have used the obvious rotational invariance of the $bc$ CFT on the cylinder.\\
Using Wick theorem (which is reviewed in Appendix \ref{sec:wick}) and in particular \eqref{Vamplitudes} we obtain
\begin{equation}
 \begin{split}
  B_{2,m}^{\tin{SFT}} \ =\ -\ 4\pi i\ \int_0^{\frac{1}{2}} ds\ s^\epsilon\ &\corr{c(i\infty)c(-i\infty)c(s)\ \V(i\infty)\ \Vaux(i\infty)}_{C_2}\ \times \\
  &\times \ \Bigg\{ \corr{j(x_2)j(x_2)}_{C_2}\ -\ m^2\ \Big( \corr{\tilde{X}(i\infty) j(0)}_{C_2} \Big)^2 \Bigg\} \ ,
 \end{split}
\end{equation}
which using  \eqref{2punti}, \eqref{ghostV} and \eqref{Vcilindro} gives
\begin{equation}
 \begin{split}
  B_{2,m}^{\tin{SFT}} \ =\ 4\ \int_0^{\frac{1}{2}} ds\ \Bigg\{ s^\epsilon\ \frac{\pi^2}{8}\ \csc^2 (\pi s) -\ m^2\ \frac{\pi^2}{2} \Bigg\} \ ,
 \end{split}
\end{equation}
where the $\epsilon$ prescription acts only on the first term because it is the only one which is  divergent.
This gives
\begin{equation}
 \begin{split}
  \int_0^{\frac{1}{2}} ds\ s^\epsilon\  \csc^2 (\pi s)\ \Bigg|_{\epsilon=0}\ & = \ \int_0^{\frac{1}{2}} ds\ \left( \csc^2 (\pi s)\ -\ \frac{1}{\pi^2 s^2} \right) \ + \int_0^{\frac{1}{2}} ds\ s^\epsilon\ \frac{1}{\pi^2 s^2} \ \Bigg|_{\epsilon=0} \ \\
  &=\ \frac{2}{\pi^2} \ +\ \frac{2^{1-\epsilon}}{\pi^2 (\epsilon - 1)} \ \Bigg|_{\epsilon=0} \ =\ 0 \ ,
 \end{split}
\end{equation}
here we have used our analytic continuation which, as explained in section \ref{sec:reg}, amounts to computing the integral in the region of the $\epsilon$-complex plane where it converges ($\textrm{Re}\,\epsilon> 1$) and then analytically continue to $\epsilon\to0$. In doing this we have also took the freedom of ignoring convergent terms proportional to $\epsilon$ since we are only interested in the $\epsilon\to 0$ limit.

Computing also the other convergent integral, we obtain again perfect match with the BCFT results
\begin{equation}
 B_{2,m}^{\tin{SFT}} \ =\ -\  m^2\ \pi^2 \ =\ B_{2,m}^{\tin{BCFT}} \ ,
\end{equation}
which leads to
\begin{equation}
  b_2 \ =\ 0 \ .
\end{equation}

\tocless\subsection{Third order}

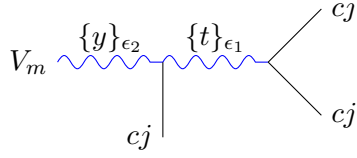
\begin{figure}[h]
 \centering
 \begin{tikzpicture}[node distance=.5cm and .5cm]
  \coordinate[label=left:$V_m$] (e1);
  \coordinate[right=1.4cm of e1] (aux1);
  \coordinate[below=1.cm of aux1,label=left:$cj$] (e2);
  \coordinate[right=1.4cm of aux1] (aux2);
  \coordinate[below right=1.cm of aux2,label=right:$cj$] (e3);
  \coordinate[above right=1.cm of aux2,label=right:$cj$] (e4);
  
  \coordinate[right=.7cm of e1,label=above:$\{y\}_{\epsilon_2}$];
  \coordinate[right=.7cm of aux1,label=above:$\{t\}_{\epsilon_1}$];

  \draw[gluon] (e1) -- (aux1);
  \draw[particle] (aux1) -- (e2);
  \draw[gluon] (aux1) -- (aux2);
  \draw[particle] (aux2) -- (e3);
  \draw[particle] (aux2) -- (e4);
 \end{tikzpicture}
 \caption{Diagram related to $\corr{V_m\hat\Psi_3 }$. The first leg of $cj$ from the left is to be understood as a commutator. The $t$ and $y$ variables are the Schwinger parameters (taking value in the interval $[0,1]$) and $\epsilon_{1,2}$ the corresponding regulators. \label{fig:ampl2}}
\end{figure}

At this level the amplitude we have to compute is
\begin{equation}
 B_{3,m}^{\tin{SFT}}\ =\ 2\pi i\ \bra{\mth{I}} V_m(i,-i) \ket{\Psi_3}  \ =\ 2\pi i\ \corr{V_m(i\infty,-i\infty) \hat\Psi_3 }_{C_3} \ ,
\end{equation}
where $\Psi_3$ is defined in \eqref{psi3}. This amplitude is depicted in Figure \ref{fig:ampl2}.

Explicitly we find
\begin{equation}
 \begin{split}
  \Psi_3\ &=\ U_4^* U_4\ \bLc \comm{ cj(0)\vk }{ \bLc\ cj\lr{\due} cj\lr{-\due} \vk }\ \\
  &=\ U_4^* U_4\ \int_0^1 dy\ y^{\epsilon_2+1}\  \int_0^{1} \frac{ dt}{2}\ t^{\epsilon_1}\  \Bigg\{ cj(y)\ j\lr{y\lr{\tfrac{t}{2}-\due}}\ j\lr{y\lr{-\tfrac{t}{2}-\due}}\  \\
  &+\ j\lr{y\lr{\tfrac{t}{2}+\due}}\ j\lr{y\lr{-\tfrac{t}{2}+\due}}\ cj(-y)\  \\
  &+\ j(y)\ \Big[ j\lr{y\lr{\tfrac{t}{2}-\due}}\ cj\lr{y\lr{-\tfrac{t}{2}-\due}}\ +\ cj\lr{y\lr{\tfrac{t}{2}-\due}}\ j\lr{y\lr{-\tfrac{t}{2}-\due}}\ \Big] \  \\
  &+\ \Big[ j\lr{y\lr{\tfrac{t}{2}+\due}}\ cj\lr{y\lr{-\tfrac{t}{2}+\due}}\ +\ cj\lr{y\lr{\tfrac{t}{2}+\due}}\ j\lr{y\lr{-\tfrac{t}{2}+\due}} \Big] \ j(-y) \Bigg\} \vk  \Bigg|_{\epsilon_1=0} \ \Bigg|_{\epsilon_2=0}\ ,
 \end{split}
\end{equation}
where $\epsilon_1$ is the regulator for the most internal propagator (the one inside the lower order contribution $\hat\Psi_2$ \eqref{psi2}) and $\epsilon_2$ is the regulator for the external propagator. From the perturbative construction of the solution, it is clear that $\epsilon_1$ should be analytically continued to zero before \\$\epsilon_2$.
Using the symmetries of the correlator in the matter and ghost sector and renaming $t/2\to t$, the whole Ellwood invariant reduces to 
\begin{equation}
 B_{3,m}^{\tin{SFT}}\ =\ 12\pi i\ \int_0^1 dy\ y^{\epsilon_2+1}\  \int_0^{\frac{1}{2}} dt\ t^{\epsilon_1}\corr{V_m(i\infty)\ cj\lr{\tfrac{3}{2}y}\ j\lr{yt}\ j\lr{-yt}  }_{C_3} \ ,
\end{equation}
It is useful to change variable with $x=\frac{3}{2}y$ and $s=ty=\frac{2}{3}xt$ so as to rewrite the integral as
\begin{equation}
 B_{3,m}^{\tin{SFT}}\ =\ 8\pi i\ \int_0^{\frac{3}{2}} dx\, x^{\epsilon_2-\epsilon_1} \int_0^{\frac{x}{3}} ds\ s^{\epsilon_1}\ \corr{V_m(i\infty)\ cj(x)\ j(s)\ j(-s)  }_{C_3} \ .
\end{equation}
Now we apply Wick theorem (see \eqref{Vamplitudes} of Appendix \ref{sec:wick}),
\begin{equation}
 \begin{split}
  B_{3,m}^{\tin{SFT}}\ &=\ 8\pi i\ \corr{c(0)\ V_m(i\infty)}_{C_3} \int_0^{\frac{3}{2}} dx\,x^{\epsilon_2-\epsilon_1} \int_0^{\frac{x}{3}} ds\ s^{\epsilon_1}\ \Bigg\{im\ \corr{\tilde{X}(i\infty) j(0)}_{C_3} \Bigg[ \corr{j(s)j(-s)}_{C_3} \ +  \\
  &+\ \corr{j(x)j(s)}_{C_3}\ +\ \corr{j(x)j(-s)}_{C_3} \Bigg] -im^3\ \Big( \corr{\tilde{X}(i\infty) j(0)}_{C_3} \Big)^3 \Bigg\} \ ,
 \end{split}
\end{equation}
and using the correlators \eqref{ghostV} and \eqref{Vcilindro}, we end up with the following integral,
\begin{equation}
 \begin{split}
  B_{3,m}^{\tin{SFT}}\ &=\ -\ 12 \int_0^{\frac{3}{2}} dx \ x^{\epsilon_2-\epsilon_1}\int_0^{\frac{x}{3}} ds\ s^{\epsilon_1}\ \Bigg\{im\ \frac{\pi\sqrt{2}}{3}\ \Bigg[ \corr{j(s)j(-s)}_{C_3} \ +  \\
  &+\ \corr{j(x)j(s)}_{C_3}\ +\ \corr{j(x)j(-s)}_{C_3} \Bigg] -im^3\ \lr{ \frac{\sqrt{2}\pi}{3}}^3 \Bigg\} \ .
 \end{split}
\end{equation}
The first integral contains a divergence in $\corr{j(s)j(-s)}$ when $s$ approaches zero. Explicitly using \eqref{2punti} this part of the amplitude is given by
\begin{equation}
 \begin{split}
  &\int_0^{\frac{3}{2}} dx\ x^{\epsilon_2-\epsilon_1}\int_0^{\frac{x}{3}} ds \ s^{\epsilon_1}\Bigg[  \corr{j(s)j(-s)}_{C_3} \ +\ \corr{j(x)j(s)}_{C_3}\ +\ \corr{j(x)j(-s)}_{C_3} \Bigg] \ \\
  &=\ \int_0^{\frac{3}{2}} dx\ x^{\epsilon_2-\epsilon_1}\int_0^{\frac{x}{3}} ds\ s^{\epsilon_1}\Bigg[  \csc^2\left[\frac{2\pi}{3}s\right]\ +\ \csc^2 \left[\frac{\pi}{3}(x+s)\right]\ +\ \csc^2 \left[\frac{\pi}{3}(x-s)\right] \Bigg] \Bigg|_{\epsilon_1=0} \ \Bigg|_{\epsilon_2=0}  \\
  &=\ \frac{3}{2 \pi} \int_0^{\frac{3}{2}} dx \ x^{\epsilon_2}\tan \left[ \frac{2\pi}{9} x \right]  \Bigg|_{\epsilon_2=0} \ =\ \frac{27}{4 \pi^2}\ \log 2 \ .
 \end{split}
\end{equation}
Notice that the integral in $x$ is convergent which tells us that we could have avoided the $\epsilon_2$ regulator. This is because the external propagator acts on a state which is in the fusion of three marginal operators and therefore it cannot contain the tachyon in its level expansion.
\begin{equation}
 \begin{split}
  B_{3,m}^{\tin{SFT}}\ &=\ i\frac{2 \sqrt{2}}{3!} m^3 \pi^3 \ -\ 3i\sqrt{2}\ m\pi\ \log 2\ \\
  &=\ B_{3,m}^{\tin{BCFT}} \ +\ (3\log 2) \ B_{1,m}^{\tin{BCFT}} \ .
 \end{split}
\end{equation}
Summarizing: from the BCFT side we found that the third order is proportional to $m^3$ \eqref{BCFTcoeff} and there are no other terms.
Instead, in the OSFT computation, at the third order we still get the same BCFT number proportional to $m^3$ but in addition to it there is another contribution coming from the peculiar renormalization implicitily defined by the propagator $\cB_0/\cL_0 $.
This is the first time in which there appears a discrepancy between the two approaches.
As a consequence the third order coefficient in the $\ls(\lambda)$ relation \eqref{comparing} is 
\begin{equation}
 b_3 \ =\ -\ 3\log 2 \ .
\end{equation}

\tocless\subsection{Fourth order}

The fourth order Ellwood invariant is given by
\begin{equation}
 B_{4,m}^{\tin{SFT}}\ =\ 2\pi i\ \bra{\mth{I}} V_m(i,-i)\ \ket{\Psi_4}  \ =\ 2\pi i\ \corr{V_m(i\infty,-i\infty)\ \hat\Psi_4 }_{C_4}\ ,
\end{equation}
there are two contributions coming from the $\Psi_{\ls}$ solution \eqref{psi4}
\begin{equation}
 \hat\Psi_4\vk \ =\ \bLc\ \left( \dcomm{cj(0)}{ \bLc\ \dcomm{cj(0)}{\hat\Psi_2}_{\tin{(2,3)}} }_{\tin{(2,4)}} \ -\ \frac{1}{2}\ \dcomm{\hat\Psi_2}{\hat\Psi_2}_{\tin{(3,3)}}  \right) \vk\ .
\end{equation}
The Ellwood invariant at this order is given by
\begin{equation}
 \begin{split}
  B_{4,m}^{\tin{SFT}}\ &=\ 2\pi i\ \Bigg( \corr{V_m(i\infty,-i\infty)\ \bLc\ \dcomm{cj(0)}{ \bLc\ \dcomm{cj(0)}{\hat\Psi_2}_{\tin{(2,3)}} }_{\tin{(2,4)}} }_{C_4}\  \\
  &\qquad\qquad\qquad\qquad -\ \frac{1}{2}\ \corr{V_m(i\infty,-i\infty)\ \bLc\ \dcomm{\hat\Psi_2}{\hat\Psi_2}_{\tin{(3,3)}} }_{C_4} \Bigg) \\
  &\equiv\ 2\pi i\ \lr{ \corr{V_m\hat{\mth{A}}_{2,4}}_{C_4}\ -\ \corr{V_m\hat{\mth{A}}_{3,3} }_{C_4} } \ .
 \end{split}
\end{equation}

\subsubsection*{First term $\corr{V_m\hat{\mth{A}}_{2,4}}$}

\begin{figure}[h]
 \centering
 \begin{tikzpicture}[node distance=.5cm and .5cm]
  \coordinate[label=left:$V_m$] (e1);
  \coordinate[right=1.4cm of e1] (aux1);
  \coordinate[below=1.cm of aux1,label=left:$cj$] (e2);
  \coordinate[right=1.4cm of aux1] (aux2);
  \coordinate[below=1.cm of aux2,label=left:$cj$] (e3);
  \coordinate[right=1.4cm of aux2] (aux3);
  \coordinate[below right=1.cm of aux3,label=right:$cj$] (e4);
  \coordinate[above right=1.cm of aux3,label=right:$cj$] (e5);
  
  \coordinate[right=.7cm of e1,label=above:$\{z\}_{\epsilon_3}$];
  \coordinate[right=.7cm of aux1,label=above:$\{y\}_{\epsilon_2}$];
  \coordinate[right=.7cm of aux2,label=above:$\{2t\}_{\epsilon_1}$];

  \draw[gluon] (e1) -- (aux1);
  \draw[particle] (aux1) -- (e2);
  \draw[gluon] (aux1) -- (aux2);
  \draw[particle] (aux2) -- (e3);
  \draw[gluon] (aux2) -- (aux3);
  \draw[particle] (aux3) -- (e4);
  \draw[particle] (aux3) -- (e5);
 \end{tikzpicture}
 \caption{First diagram $\corr{V_m\hat{\mth{A}}_{2,4}}$. The first two legs of $cj$ from the left are to be understood as commutators. The $z$, $y$ and $2t$  variables are the Schwinger parameters (taking value in the interval $[0,1]$). The integration variable  $s$  in \eqref{quart-1} is related to the Schwinger parameters as $s=t y$ .  \label{fig:ampl41}}
\end{figure}
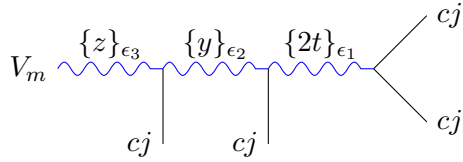

In the first term, as before, we need to compute the commutator of the insertions and apply the propagators,
\begin{equation}
 \hat{\mth{A}}_{2,4} \vk\ =\ U_5^* U_5\ \bLc\ \dcomm{cj(0)}{ \bLc\ \dcomm{cj(0)}{\hat\Psi_2}_{\tin{(2,3)}} }_{\tin{(2,4)}} \vk \ .
\end{equation}
The corresponding amplitude is depicted in Figure \ref{fig:ampl41}.

Applying the two propagators, the amplitude takes the form
\begin{equation}
 \begin{split}
  &\corr{V_m\hat{\mth{A}}_{2,4}}_{C_4}\ =\ -\ 3\int_0^1 dz\ z^{2+\epsilon_3}\int_0^1 dy\ y^{\epsilon_2-\epsilon_1} \int_0^{\frac{y}{2}} ds\ s^{\epsilon_1} \ \Bigg\langle\ V_m(i\infty)\ \times \\
  &\times\ \Bigg\{ j\lr{\tfrac{3}{2}z}\ j\lr{z\lr{y-\due}}\ j\lr{z\lr{s-\due y-\due}}\ j\lr{z\lr{-s-\due y-\due}}  \times \\
  &\times\ \Big[ c\lr{\tfrac{3}{2}z}\ +\ c\lr{z\lr{y-\due}}\ +\ c\lr{z\lr{s-\due y-\due}}\ +\ c\lr{z\lr{-s-\due y-\due}} \Big] \  \\
  &+\ j\lr{\tfrac{3}{2}z}\ j\lr{z\lr{-y-\due}}\ j\lr{z\lr{s+\due y-\due}}\ j\lr{z\lr{-s+\due y-\due}}  \times \\
  &\times\Big[ c\lr{\tfrac{3}{2}z}\ +\ c\lr{z\lr{-y-\due}}\ +\ c\lr{z\lr{s+\due y-\due}}\ +\ c\lr{z\lr{-s+\due y-\due}} \Big] \Bigg\} \ \Bigg\rangle_{C_4} \Bigg|_{\epsilon_1=0} \ \Bigg|_{\epsilon_2=0} \ \Bigg|_{\epsilon_3=0} .
 \end{split}\label{quart-1}
\end{equation}
Using the symmetries of the problem translating the correlators $\xi \to \xi +\frac{z}{2}$ and changing variables $w=sz$ and $x=yz$, the amplitude simplifies 
\begin{equation}
 \begin{split}
  \corr{V_m\hat{\mth{A}}_{2,4}}_{C_4}\ =\ -\ &6\int_0^2 dz\ z^{\epsilon_3+\epsilon_2-2\epsilon_1} \int_0^{\frac{z}{2}} dx \int_0^{\frac{x}{2}} dw\ w^{\epsilon_1}\ \times \\
  &\times\ \Bigg\{ \corr{ V_m(i\infty)\ j(z)\ j(x)\ j\lr{w-\tfrac{x}{2}}\ j\lr{-w-\tfrac{x}{2}} }_{C_4} \  \\
  &\qquad +\ \corr{ V_m(i\infty)\ j(z)\ j\lr{w+\tfrac{x}{2}}\ j\lr{-w+\tfrac{x}{2}}\ j(-x) }_{C_4} \ \Bigg\} \ .
 \end{split}
 \label{corr41}
\end{equation}

\subsubsection*{Second term $\corr{V_m\hat{\mth{A}}_{3,3} }$}

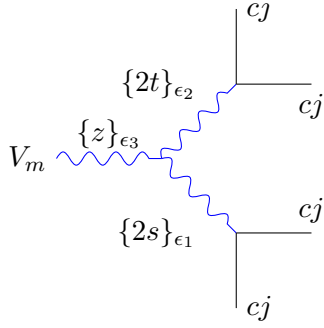
\begin{figure}[h]
 \centering
 \begin{tikzpicture}[node distance=.5cm and .5cm]
  \coordinate[label=left:$V_m$] (e1);
  \coordinate[right=1.4cm of e1] (aux1);
  \coordinate[above right=1.4cm of aux1] (aux2);
  \coordinate[below right=1.4cm of aux1] (aux3);
  \coordinate[right=1.cm of aux2,label=below:$cj$] (e2);
  \coordinate[above=1.cm of aux2,label=right:$cj$] (e3);
  \coordinate[below=1.cm of aux3,label=right:$cj$] (e4);
  \coordinate[right=1.cm of aux3,label=above:$cj$] (e5);
  
  \coordinate[right=.7cm of e1,label=above:$\{z\}_{\epsilon_3}$];
  \coordinate[left=.4cm of aux2,label=left:$\{2t\}_{\epsilon_2}$];
  \coordinate[left=.4cm of aux3,label=left:$\{2s\}_{\epsilon_1}$];

  \draw[gluon] (e1) -- (aux1);
  \draw[gluon] (aux1) -- (aux2);
  \draw[gluon] (aux1) -- (aux3);
  \draw[particle] (aux2) -- (e2);
  \draw[particle] (aux2) -- (e3);
  \draw[particle] (aux3) -- (e4);
  \draw[particle] (aux3) -- (e5);
 \end{tikzpicture}
 \caption{Second diagram $\corr{V_m\hat{\mth{A}}_{3,3} }$. $z$, $2t$ and $2s$ are the Schwinger parameters (taking value in the interval $[0,1]$). \label{fig:ampl42}}
\end{figure}

The second term is the Ellwood invariant of $\hat{\mth{A}}_{3,3}$,
\begin{equation}
 \hat{\mth{A}}_{3,3}\vk\ =\ \frac{1}{2}\ U_5^* U_5\ \bLc\ \dcomm{\hat\Psi_2}{\hat\Psi_2}_{\tin{(3,3)}} \vk\ ,
\end{equation}
and it is depicted in Figure \ref{fig:ampl42}.

Explicitly we have to compute the star product of two $\Psi_2$ and then act with a propagator $\cB_0/\cL_0$:
\begin{equation}
 \begin{split}
  \corr{V_m\hat{\mth{A}}_{3,3} }_{C_4}\ =\ & \int_0^1 dz\ z^{2+\epsilon_3}\ \int_0^\frac{1}{2} dt\ t^{\epsilon_2}\int_0^{\frac{1}{2}} ds\ s^{\epsilon_1}  \times\\
  &\times\ 2\langle\ V_m(i\infty)\ j(z(t+1))\ j(z(-t+1))\ j(z(s-1))\ j(z(-s-1))\ \times\\
  &\qquad \times\ \Big[ c(z(t+1))\ +\ c(z(-t+1))\ +\ c(z(s-1))\ +\ c(z(-s-1))  \Big] \ \rangle_{C_4} \Bigg|_{\epsilon_{1,2}=0} \  \Bigg|_{\epsilon_3=0} \ .
 \end{split}
\end{equation}
Again the four different insertions of the ghosts contribute in the same way, and therefore
\begin{equation}
 \begin{split}
  \corr{V_m\hat{\mth{A}}_{3,3} }_{C_4}\ =\ & 8\int_0^1 dz\ z^{2+\epsilon_3}\ \int_0^\frac{1}{2} dt\ t^{\epsilon_2}\int_0^{\frac{1}{2}} ds\ s^{\epsilon_1} \ \times\\
  & \times\ \langle\ V_m(i\infty)\ cj(z(t+1))\ j(z(-t+1))\ j(z(s-1))\ j(z(-s-1))\ \rangle_{C_4} \ .
 \end{split}
\end{equation}
With the change of variable $x=zt$ and $y=zs$,
\begin{equation}
 \begin{split}
  \corr{V_m\hat{\mth{A}}_{3,3} }_{C_4}\ =\ & 8\int_0^1 dz\  z^{\epsilon_3-\epsilon_2-\epsilon_1}\int_0^{\frac{z}{2}} dx\ x^{\epsilon_2} \int_0^{\frac{z}{2}} dy\ y^{\epsilon_1}\ \times\\
  &\qquad \times\ \langle\ V_m(i\infty)\ cj(z+x)\ j(z-x)\ j(-z+y))\ j(-z-y)\ \rangle_{C_4} \ .
 \end{split}
 \label{corr42}
\end{equation}

\subsubsection*{Complete $B_{4,m}^{\tin{SFT}}$}

The complete term at this order is given by summing the two integrals \eqref{corr41} and \eqref{corr42},
\begin{equation}
 \begin{split}
 & B_{4,m}^{\tin{SFT}} = - 12\pi i  \int_0^2 dz\ z^{\epsilon_3+\epsilon_2-2\epsilon_1} \int_0^{\frac{z}{2}} dx \int_0^{\frac{x}{2}} dw\ w^{\epsilon_1}\Bigg\{ \corr{ V_m(i\infty)\ j(z) j(x) j\lr{w-\tfrac{x}{2}} j\lr{-w-\tfrac{x}{2}} }_{C_4}  \\
  &\hspace{5.5cm} +\ \corr{ V_m(i\infty)\ j(z)\ j\lr{w+\tfrac{x}{2}}\ j\lr{-w+\tfrac{x}{2}}\ j(-x) }_{C_4} \ \Bigg\} \\
  &-16\pi i \int_0^1 dz\  z^{\epsilon_3-\epsilon_2-\epsilon_1}\int_0^{\frac{z}{2}} dx\ x^{\epsilon_2} \int_0^{\frac{z}{2}} dy\ y^{\epsilon_1}\ \langle\ V_m(i\infty)\ cj(z+x)\ j(z-x)\ j(-z+y))\ j(-z-y)\ \rangle_{C_4} \ .
 \end{split}
\end{equation}
As in the previous order $\ls^3$, here also we have contribution from three different powers of the winding number $m$ coming from the different contractions in Wick theorem \eqref{Vamplitudes}. This means that we can write $B_{4,m}^{\tin{SFT}}$ in terms of the $B_{k,m}^{\tin{BCFT}} \sim m^k $,
\begin{equation}
 \begin{split}
  B_{4,m}^{\tin{SFT}}\ =\ a_0\ B_{0,m}^{\tin{BCFT}}\  +\  a_2\ B_{2,m}^{\tin{BCFT}} \ +\ a_4\ B_{4,m}^{\tin{BCFT}} \ .
 \end{split}
\end{equation}
From this consideration (setting to zero all the regulators as they are not important here) we find that
\begin{equation}
 \begin{split}
  a_4\ B_{4,m}^{\tin{BCFT}}\ &=\ \frac{6}{8}\ m^4 \pi^4 \int_0^2 dz \int_0^{\frac{z}{2}} dx \int_0^{\frac{x}{2}} dw\ +\ \frac{1}{2}\ m^4 \pi^4 \int_0^1 dz \int_0^{\frac{z}{2}} dx \int_0^{\frac{z}{2}} dw\ \\
  &=\ \frac{m^4 \pi^4}{6}\ .
 \end{split}
\end{equation}
\begin{equation}
 \begin{split}
  a_2\ B_{2,m}^{\tin{BCFT}} =\ -\ & \frac{3}{16}\ m^2 \pi^4  \int_0^2 dz\ z^{\epsilon_3+\epsilon_2-2\epsilon_1} \int_0^{\frac{z}{2}} dx \int_0^{\frac{x}{2}} dw\ w^{\epsilon_1}\ \Bigg\{ 2 \csc^2 \left[ \frac{\pi}{2} w \right] \ +\ 2\ \csc^2 \left[ \frac{\pi}{4} \lr{w-\frac{3}{2}x} \right]  \\
  &+\ 2\ \csc^2 \left[ \frac{\pi}{4} (w+\frac{3}{2}x) \right] \ +\ \csc^2 \left[ \frac{\pi}{4} (z+x) \right] \ +\ \csc^2 \left[ \frac{\pi}{4} (z-x) \right] \\
  &+\ \csc^2 \left[ \frac{\pi}{4} \lr{z+w+\frac{x}{2}} \right] \ +\ \csc^2 \left[ \frac{\pi}{4} \lr{z+w-\frac{x}{2}} \right] \\
  &+\ \csc^2 \left[ \frac{\pi}{4} \lr{z-w+\frac{x}{2}} \right] \ +\ \csc^2 \left[ \frac{\pi}{4} \lr{z-w-\frac{x}{2}} \right] \Bigg\} \   \Bigg|_{\epsilon_1=0} \ \Bigg|_{\epsilon_2=0} \ \Bigg|_{\epsilon_3=0}\\
  -\ \frac{1}{4}&\ m^2 \pi^4  \int_0^1 dz\  z^{\epsilon_3-\epsilon_2-\epsilon_1}\int_0^{\frac{z}{2}} dx\ x^{\epsilon_2} \int_0^{\frac{z}{2}} dw\ w^{\epsilon_1} \Bigg\{ \csc^2 \left[ \frac{\pi}{2} w \right] \ + \csc^2 \left[ \frac{\pi}{2} x \right]  \\
  &+\ \csc^2 \left[ \frac{\pi}{4} \lr{w+2z+x} \right] \ +\ \csc^2 \left[ \frac{\pi}{4} \lr{w+2z-x} \right] \\
  &+\ \csc^2 \left[ \frac{\pi}{4} \lr{w-2z+x} \right] \ +\ \csc^2 \left[ \frac{\pi}{4} \lr{-w+2z+x} \right] \Bigg\} \ \Bigg|_{\epsilon_{1,2}=0} \  \Bigg|_{\epsilon_3=0} \\
  =\ &\big( 2\ m^2 \pi^2\big) \ 3\ \log 2 \ .
 \end{split}
\end{equation}
\begin{equation}
 \begin{split}
  a_0\ B_{0,m}^{\tin{BCFT}}\ =\ -\ & \frac{3}{32}\ \pi^4  \int_0^2 dz\ z^{\epsilon_3+\epsilon_2-2\epsilon_1} \int_0^{\frac{z}{2}} dx \int_0^{\frac{x}{2}} dw\ w^{\epsilon_1}\ \Bigg\{ \csc^2 \left[ \frac{\pi}{2} w \right] \ \csc^2 \left[ \frac{\pi}{4} (z+x) \right]  \\
  &+ \csc^2 \left[ \frac{\pi}{2} w \right] \ \csc^2 \left[ \frac{\pi}{4} (z-x) \right]  \\
  &+\ \csc^2 \left[ \frac{\pi}{4} \lr{w+\frac{3}{2}x} \right] \ \csc^2 \left[ \frac{\pi}{4} \lr{z+w-\frac{x}{2}} \right]  \\
  &+\ \csc^2 \left[ \frac{\pi}{4} \lr{w+\frac{3}{2}x} \right] \ \csc^2 \left[ \frac{\pi}{4} \lr{z-w+\frac{x}{2}} \right]  \\
  &+\ \csc^2 \left[ \frac{\pi}{4} \lr{w-\frac{3}{2}x} \right] \ \csc^2 \left[ \frac{\pi}{4} \lr{z-w-\frac{x}{2}} \right]  \\
  &+\ \csc^2 \left[ \frac{\pi}{4} \lr{w-\frac{3}{2}x} \right] \ \csc^2 \left[ \frac{\pi}{4} \lr{z+w+\frac{x}{2}} \right] \Bigg\} \   \Bigg|_{\epsilon_1=0} \ \Bigg|_{\epsilon_2=0} \ \Bigg|_{\epsilon_3=0}\\
  -\ \frac{1}{4}&\ m^2 \pi^4  \int_0^1 dz\  z^{\epsilon_3-\epsilon_2-\epsilon_1}\int_0^{\frac{z}{2}} dx\ x^{\epsilon_2} \int_0^{\frac{z}{2}} dw\ w^{\epsilon_1} \Bigg\{  \csc^2 \left[ \frac{\pi}{2} w \right] \ \csc^2 \left[ \frac{\pi}{2} x \right]  \\
  &+\ \csc^2 \left[ \frac{\pi}{4} \lr{w+2z+x} \right] \ \csc^2 \left[ \frac{\pi}{4} \lr{w-2z+x} \right] \\
  &+\ \csc^2 \left[ \frac{\pi}{4} \lr{w+2z-x} \right] \ \csc^2 \left[ \frac{\pi}{4} \lr{-w+2z+x} \right] \Bigg\} \Bigg|_{\epsilon_{1,2}=0} \  \Bigg|_{\epsilon_3=0} \\
  =\ &0 \ .
  \end{split}
\end{equation}

Using these results we find that
\begin{equation}
 B_{4,m}^{\tin{SFT}}\ =\ B_{4,m}^{\tin{BCFT}}\ +\ \lr{ 6 \log 2}\ B_{2,m}^{\tin{BCFT}}\ ,
\end{equation}
which corresponds to
\begin{equation}
 b_4\ =\ 0 \ .
\end{equation}

\tocless\subsection{Fifth order}

At the fifth order the solution is composed of three terms,
\begin{equation}
 \begin{split}
  \hat\Psi_5\vk \ =\ \bLc\ \Bigg( & -\ \dcomm{cj(0)}{ \bLc\ \dcomm{ cj(0) }{ \bLc\ \dcomm{cj(0)}{\hat\Psi_2}_{\tin{(2,3)}} }_{\tin{(2,4)}} }_{\tin{(2,5)}} \  \\
  &-\ \dcomm{\Psi_2}{ \bLc\ \dcomm{cj(0)}{\hat\Psi_2}_{\tin{(2,3)}} }_{\tin{(3,4)}} \ +\ \frac{1}{2}\ \dcomm{cj(0)}{ \bLc\ \dcomm{\hat\Psi_2}{\hat\Psi_2}_{\tin{(3,3)}} }_{\tin{(2,5)}} \Bigg) \vk\ \ .
 \end{split}
\end{equation}
Then the Ellwood invariant we have to compute is
\begin{equation}
 \begin{split}
  B_{5,m}^{\tin{SFT}}\ &=\ 2\pi i\ \bra{\mth{I}} V_m(i,-i)\ \ket{\Psi_5} \ =\ 2\pi i\ \corr{V_m(i\infty,-i\infty)\ \hat\Psi_5 }_{C_5}\ \\
  &=\ 2\pi i\ \Bigg( -\ \corr{V_m(i\infty,-i\infty)\ \bLc\ \dcomm{cj(0)}{ \bLc\ \dcomm{ cj(0) }{ \bLc\ \dcomm{cj(0)}{\hat\Psi_2}_{\tin{(2,3)}} }_{\tin{(2,4)}} }_{\tin{(2,5)}}  }_{C_5}\  \\
  &\qquad\qquad\qquad\qquad -\ \corr{V_m(i\infty,-i\infty)\ \bLc\ \dcomm{\Psi_2}{ \bLc\ \dcomm{cj(0)}{\hat\Psi_2}_{\tin{(2,3)}} }_{\tin{(3,4)}} }_{C_5} \  \\
  &\qquad\qquad\qquad\qquad +\ \frac{1}{2}\ \corr{V_m(i\infty,-i\infty)\ \bLc\ \dcomm{cj(0)}{ \bLc\ \dcomm{\hat\Psi_2}{\hat\Psi_2}_{\tin{(3,3)}} }_{\tin{(2,5)}} }_{C_5} \Bigg) \ \\
  &\equiv\ 2\pi i\ \lr{ -\ \corr{V_m\hat{\mth{A}}_{2,5} }_{C_5} \ -\ \corr{V_m\hat{\mth{A}}_{3,4} }_{C_5} \ + \ \frac{1}{2}\ \corr{V_m\hat{\mth{A}}^{3,3}_{2,5} }_{C_5}  } \ .
 \end{split}
\end{equation}

\subsubsection*{First term $\corr{V_m\hat{\mth{A}}_{2,5} }$}

The first term involves the state
\begin{equation}
 \hat{\mth{A}}_{2,5} \vk\ =\ U_6^* U_6\ \bLc\ \dcomm{cj(0)}{ \bLc\ \dcomm{ cj(0) }{ \bLc\ \dcomm{cj(0)}{\hat\Psi_2}_{\tin{(2,3)}} }_{\tin{(2,4)}} }_{\tin{(2,5)}}  \vk \ .
\end{equation}
\begin{figure}[h]
 \centering
 \begin{tikzpicture}[node distance=.5cm and .5cm]
  \coordinate[label=left:$V_m$] (e1);
  \coordinate[right=1.4cm of e1] (aux1);
  \coordinate[below=1.cm of aux1,label=left:$cj$] (e2);
  \coordinate[right=1.4cm of aux1] (aux2);
  \coordinate[below=1.cm of aux2,label=left:$cj$] (e3);
  \coordinate[right=1.4cm of aux2] (aux3);
  \coordinate[below=1.cm of aux3,label=left:$cj$] (e4);
  \coordinate[right=1.4cm of aux3] (aux4);
  \coordinate[below right=1.cm of aux4,label=right:$cj$] (e5);
  \coordinate[above right=1.cm of aux4,label=right:$cj$] (e6);
  
  \coordinate[right=.7cm of e1,label=above:$\{t_4\}_{\epsilon_4}$];
  \coordinate[left=.7cm of aux2,label=above:$\{t_3\}_{\epsilon_3}$];
  \coordinate[left=.7cm of aux3,label=above:$\{t_2\}_{\epsilon_2}$];
  \coordinate[left=.7cm of aux4,label=above:$\{2t_1\}_{\epsilon_1}$];

  \draw[gluon] (e1) -- (aux1);
  \draw[particle] (aux1) -- (e2);
  \draw[gluon] (aux1) -- (aux2);
  \draw[particle] (aux2) -- (e3);
  \draw[gluon] (aux2) -- (aux3);
  \draw[particle] (aux3) -- (e4);
  \draw[gluon] (aux3) -- (aux4);
  \draw[particle] (aux4) -- (e5);
  \draw[particle] (aux4) -- (e6);
 \end{tikzpicture}
 \caption{First diagram $\corr{V_m\hat{\mth{A}}_{2,5} }$.  \label{fig:ampl51}}
\end{figure}

The amplitude to compute is depicted in Figure \ref{fig:ampl51} and after some manipulations involving changes of variables and conformal transformations we get
\begin{equation}
 \begin{split}
  2\pi i\ &\corr{V_m\hat{\mth{A}}_{2,5} }_{C_5} \ =\ -\ 48\pi i \int_0^{\frac{5}{2}} dT\ T^{\epsilon_4-\epsilon_3-\epsilon_2+\epsilon_1} \int_0^{\frac{T}{5}} dX\ X^{\epsilon_3+\epsilon_2-2\epsilon_1} \int_0^{2X} dY \int_0^{\frac{Y}{2}} dZ\ Z^{\epsilon_1}\ \times \\
  \times\ &\Bigg\{ \corr{ V_m(i\infty)\ cj(T)\ j(3X)\ j(Y-X)\ j\lr{Z-\tfrac{Y}{2}-X}\ j\lr{-Z-\tfrac{Y}{2}-X} }_{C_5} \\
  &+\ \corr{ V_m(i\infty)\ cj(T)\ j(3X)\ j\lr{Z+\tfrac{Y}{2}-X} \ j\lr{-Z+\tfrac{Y}{2}-X}\ j(-Y-X) }_{C_5} \\
  &+\ \corr{ V_m(i\infty)\ cj(T)\ j\lr{Z+\tfrac{Y}{2}+X}\ j\lr{-Z+\tfrac{Y}{2}+X}\ j(-Y+X)\ j\lr{-3X} }_{C_5} \\
  &+\ \corr{ V_m(i\infty)\ cj(T)\ j(Y+X)\ j\lr{Z-\tfrac{Y}{2}+X}\ j\lr{-Z-\tfrac{Y}{2}+X}\ j\lr{-3X} }_{C_5} \Bigg\} \Bigg|_{\epsilon_1=0}  \Bigg|_{\epsilon_2=0}  \Bigg|_{\epsilon_3=0}  \Bigg|_{\epsilon_4=0} \ ,
 \end{split}
\end{equation}
where the Schwinger parameters $t_1,t_2,t_3,t_4$ are related to the integration variables as
\begin{equation}
 \begin{split}
  T\ &=\ t_4 \ , \\
  X\ &=\ t_4 t_3 \ , \\
  Y\ &=\ t_4 t_3 t_2 \ , \\
  Z\ &=\ t_4 t_3 t_2 t_1 \ .
 \end{split}
\end{equation}
\\

\subsubsection*{Second term $\corr{V_m\hat{\mth{A}}_{3,4} }$}

The state here is
\begin{equation}
 \hat{\mth{A}}_{3,4} \vk\ =\ U_6^* U_6\ \bLc\ \dcomm{\Psi_2}{ \bLc\ \dcomm{cj(0)}{\hat\Psi_2}_{\tin{(2,3)}} }_{\tin{(3,4)}} \vk \ ,
\end{equation}
\begin{figure}[h!]
\centering
 \begin{tikzpicture}[node distance=.5cm and .5cm]
  \coordinate[label=left:$V_m$] (e1);
  \coordinate[right=1.4cm of e1] (aux1);
  \coordinate[right=1.4cm of aux1] (aux2);
  \coordinate[below=1.4cm of aux1] (aux3);
  \coordinate[below right=1.cm of aux3,label=below:$cj$] (e3);
  \coordinate[below left=1.cm of aux3,label=below:$cj$] (e4);
  \coordinate[right=1.4cm of aux2] (aux4);
  \coordinate[below=1.cm of aux2,label=left:$cj$] (e5);
  \coordinate[below right=1.cm of aux4,label=right:$cj$] (e6);
  \coordinate[above right=1.cm of aux4,label=right:$cj$] (e7);
  
  \coordinate[right=.7cm of e1,label=above:$\{t_4\}_{\epsilon_4}$];
  \coordinate[left=.7cm of aux2,label=above:$\{t_3\}_{\epsilon_3}$];
  \coordinate[below=.7cm of aux1,label=left:$\{2t_1\}_{\epsilon_1}$];
  \coordinate[left=.7cm of aux4,label=above:$\{2t_2\}_{\epsilon_2}$];

  \draw[gluon] (e1) -- (aux1);
  \draw[gluon] (aux1) -- (aux3);
  \draw[gluon] (aux1) -- (aux2);
  \draw[particle] (aux3) -- (e3);
  \draw[particle] (aux3) -- (e4);
  \draw[particle] (aux2) -- (e5);
  \draw[gluon] (aux2) -- (aux4);
  \draw[particle] (aux4) -- (e6);
  \draw[particle] (aux4) -- (e7);
  \end{tikzpicture}
 \caption{Second diagram $\corr{V_m\hat{\mth{A}}_{3,4} }$. }
\end{figure}
and its Ellwood invariant is
\begin{equation}
 \begin{split}
  2\pi i\ &\corr{V_m\hat{\mth{A}}_{3,4} }_{C_5}\ =\ -\ 24\pi i \int_0^{\frac{5}{2}} dT\ T^{\epsilon_4-\epsilon_3-\epsilon_2} \int_0^{\frac{2}{5}T} dX\ X^{\epsilon_3-\epsilon_2} \int_0^{\frac{X}{2}} dY\ Y^{\epsilon_2} \int_0^{\frac{T}{2}} dZ\ Z^{\epsilon_1} \times \\
  \times\ &\Bigg\{ \corr{ V_m(i\infty)\ cj(T+Z)\ j(T-Z)\ j(X)\ j\lr{Y-\tfrac{X}{2}}\ j\lr{-Y-\tfrac{X}{2}} }_{C_5}\  \\
  &+\ \corr{ V_m(i\infty)\ cj(T+Z)\ j(T-Z)\ j\lr{Y+\tfrac{X}{2}}\ j\lr{-Y+\tfrac{X}{2}}\ j(-X) }_{C_5} \Bigg\}  \Bigg|_{\epsilon_1=0}  \Bigg|_{\epsilon_2=0}  \Bigg|_{\epsilon_3=0}  \Bigg|_{\epsilon_4=0} \ ,
 \end{split}
\end{equation}
where the integration variables are related to the Schwinger parameters as
\begin{equation}
 \begin{split}
  T\ &=\ t_4 \ , \\
  X\ &=\ t_4 t_3 \ , \\
  Y\ &=\ t_4 t_3 t_2 \ , \\
  Z\ &=\ t_4 t_1 \ .
 \end{split}
\end{equation}

\subsubsection*{Third term $\corr{V_m\hat{\mth{A}}^{3,3}_{2,5} }$}

The last term involves the state
\begin{equation}
 \hat{\mth{A}}^{3,3}_{2,5}\vk\ =\ \frac{1}{2}\ U_5^* U_5\ \bLc\ \dcomm{cj(0)}{\dcomm{\hat\Psi_2}{\hat\Psi_2}_{\tin{(3,3)}}}_{\tin{(2,5)}} \vk \ ,
\end{equation}
\begin{figure}[h]
 \centering
 \begin{tikzpicture}[node distance=.5cm and .5cm]
  \coordinate[label=left:$V_m$] (e1);
  \coordinate[right=1.4cm of e1] (aux1);
  \coordinate[below=1.cm of aux1,label=left:$cj$] (e2);
  \coordinate[right=1.4cm of aux1] (aux2);
  \coordinate[above right=1.4cm of aux2] (aux3);
  \coordinate[below right=1.4cm of aux2] (aux4);
  \coordinate[right=1.cm of aux3,label=below:$cj$] (e3);
  \coordinate[above=1.cm of aux3,label=right:$cj$] (e4);
  \coordinate[below=1.cm of aux4,label=right:$cj$] (e5);
  \coordinate[right=1.cm of aux4,label=above:$cj$] (e6);
  
  \coordinate[right=.7cm of e1,label=above:$\{t_4\}_{\epsilon_4}$];
  \coordinate[left=.7cm of aux2,label=above:$\{t_3\}_{\epsilon_3}$];
  \coordinate[below left=.7cm of aux3,label=above left:$\{2t_2\}_{\epsilon_2}$];
  \coordinate[above left=.7cm of aux4,label=below left:$\{2t_1\}_{\epsilon_1}$];

  \draw[gluon] (e1) -- (aux1);
  \draw[particle] (aux1) -- (e2);
  \draw[gluon] (aux1) -- (aux2);
  \draw[gluon] (aux2) -- (aux3);
  \draw[gluon] (aux2) -- (aux4);
  \draw[particle] (aux3) -- (e3);
  \draw[particle] (aux3) -- (e4);
  \draw[particle] (aux4) -- (e5);
  \draw[particle] (aux4) -- (e6);
 \end{tikzpicture}
 \caption{Third diagram $\corr{V_m\hat{\mth{A}}^{3,3}_{2,5} }$. }
\end{figure}
and it becomes
\begin{equation}
 \begin{split}
  2\pi i\ &\corr{V_m\hat{\mth{A}}^{3,3}_{2,5} }_{C_5}\ =\ 32\pi i \int_0^{\frac{5}{2}} dT\ T^{\epsilon_4-\epsilon_3} \int_0^{\frac{2}{5}T} dX\ X^{\epsilon_3-\epsilon_2-\epsilon_1} \int_0^{\frac{X}{2}} dY\ Y^{\epsilon_2} \int_0^{\frac{X}{2}} dZ\ Z^{\epsilon_1}\ \times \\
  \times\ &\Bigg\{ \corr{ V_m(i\infty)\ cj(T)\ j(X+Y)\ j(X-Y)\ j\lr{Z-X}\ j\lr{-Z-X} }_{C_5} \\
  &+\ \corr{ V_m(i\infty)\ cj(X+Y)\ j(X-Y)\ j\lr{Z-X}\ j\lr{-Z-X}\ j(-T) }_{C_5} \Bigg\}  \Bigg|_{\epsilon_1=0}  \Bigg|_{\epsilon_2=0}  \Bigg|_{\epsilon_3=0}  \Bigg|_{\epsilon_4=0} \ ,
 \end{split}
\end{equation}
where the Schwinger parameters are related to the integration variables as
\begin{equation}
 \begin{split}
  T\ &=\ t_4 \ , \\
  X\ &=\ t_4 t_3 \ , \\
  Y\ &=\ t_4 t_3 t_2 \ , \\
  Z\ &=\ t_4 t_3 t_1 \ .
 \end{split}
\end{equation}

\subsubsection*{Complete $B_{5,m}^{\tin{SFT}}$}

Now we have to compute the three Ellwood invariants as we have done in previous examples. Again the total Ellwood invariant can be expanded in terms of the first, third and fifth power of the winding number,
\begin{equation}
 B_{5,m}^{\tin{SFT}}\ =\ a_1\ B_{1,m}^{\tin{BCFT}}\  +\  a_3\ B_{3,m}^{\tin{BCFT}} \ +\ a_5\ B_5^{\tin{BCFT}} \ .\label{tot-five}
\end{equation}
The $a_5$ coefficient is easily computed
\begin{equation}
 \begin{split}
  a_5\ B_5^{\tin{BCFT}}\ =\ & -\ 480\ \frac{(i m)^5}{5!}\ \Big( \corr{X(i\infty) j(0)}_{C_5} \Big)^5 \int_0^{\frac{5}{2}} dT \int_0^{\frac{T}{5}} dX \int_0^{2X} dY \int_0^{\frac{Y}{2}} dZ\  \\
  &-\ 300\ \frac{(i m)^5}{5!}\ \Big( \corr{X(i\infty) j(0)}_{C_5} \Big)^5  \int_0^{\frac{5}{2}} dT \int_0^{\frac{2}{5}T} dX \int_0^{\frac{X}{2}} dY \int_0^{\frac{T}{2}} dZ\  \\
  &-\ 80\ \frac{(i m)^5}{5!}\ \Big( \corr{X(i\infty) j(0)}_{C_5} \Big)^5  \int_0^{\frac{5}{2}} dT \int_0^{\frac{2}{5}T} dX \int_0^{\frac{X}{2}} dY \int_0^{\frac{X}{2}} dZ\ \\
  =\ &i \frac{4\sqrt{2}}{5!}\ \pi^5 m^5 \ = \ B_5^{\tin{BCFT}} \ ,
 \end{split}
\end{equation}
which gives the usual exact match with the BCFT results.
As far as $a_3$ is concerned the computation follows closely the fourth order (with one more integral) and everything can be analitically done giving the result
\begin{equation}
 a_3 \ =\ 9 \log 2\ .
\end{equation}
This is precisely the needed number to ensure that $b_5$ is $m$-independent \eqref{b5} so this is a consistency check. 

Let us now address the $a_1$ coefficient, which is determined by the $O(m)$ winding number contribution in \eqref{tot-five}.  This is generated by the term from the Wick theorem with the maximal number of contractions between the $j$'s and computing the four dimensional integrals (coming from the three diagrams) analitically is not possible. Therefore we procede analitically as far as we can and then we resort to numerics.
The $Y$ and $Z$ integrals can be analitically computed in all of the three diagrams, including the subtraction of the tachyon divergence.
\begin{figure}[h!]
   \centering
   \includegraphics[scale=.7]{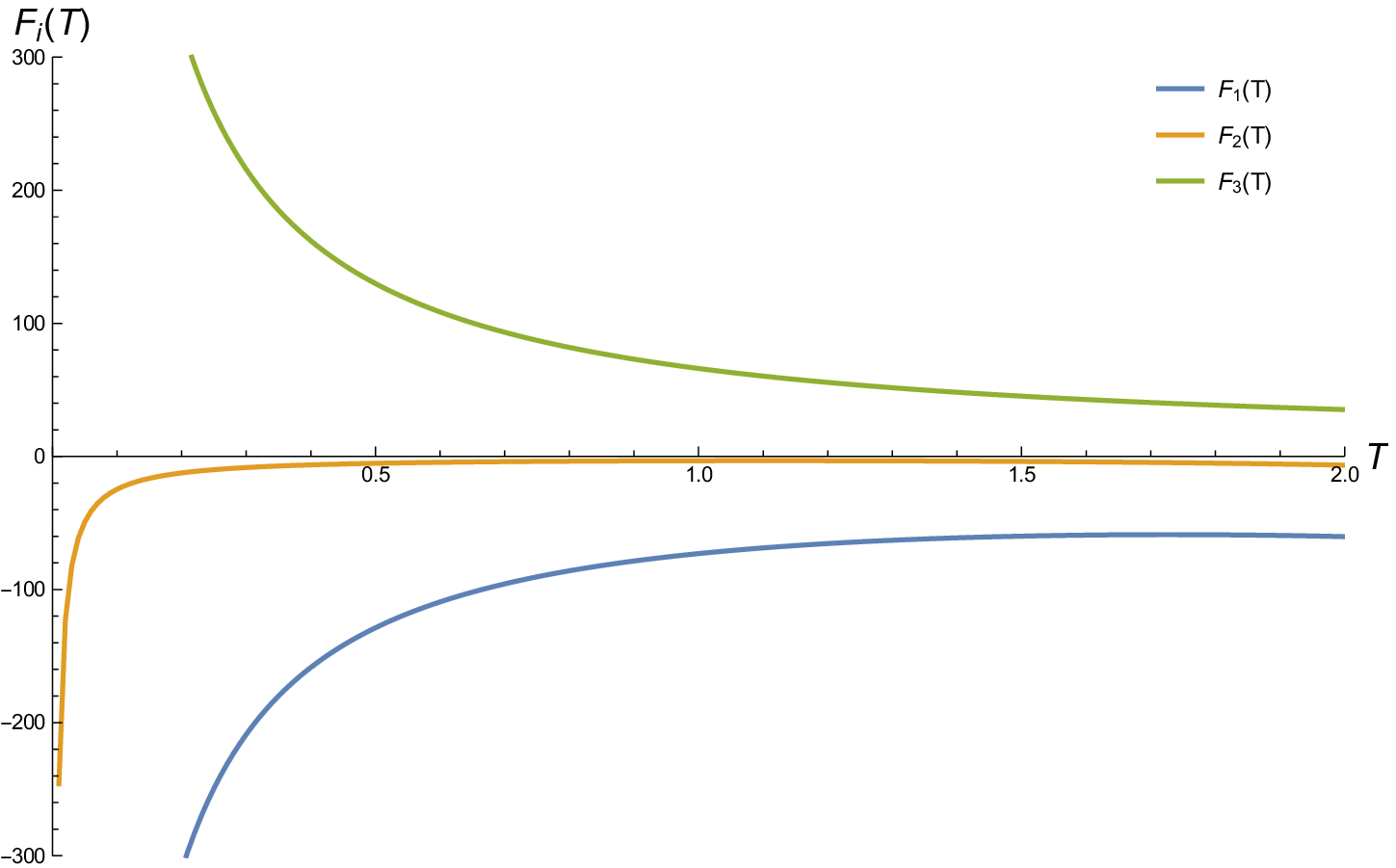}
   \caption{ The three diagrams are divergent with a simple pole in $T=0$. \label{fig:div1} }
\end{figure}

Rescaling the $X$ variable in the first diagram $X\rightarrow 2X$, the $O(m)$ contribution $E_i$ from each  diagram  is reduced to an expression of the form
\begin{eqnarray}
a_1 \ B_1^{\tin{BCFT}}&=&E_1+E_2+E_3,\quad\quad O(m)\nonumber\\
 E_i\ &=&\ \int_0^{\frac{5}{2}} dT\ T^{\epsilon_4-\epsilon_3} \int_0^{\frac{2}{5}T} dX\ X^{\epsilon_3}\ f_i(T,X)\ \Bigg|_{\epsilon_3=0}  \Bigg|_{\epsilon_4=0} , \quad i=1,2,3 \ ,
\end{eqnarray}
where the function $f_i$ are known analitically.
To renormalize the tachyon divergence in the $X$ integration we explicitly subtract the second order pole in $X$ from the function $f_i$ in the following way
\begin{equation}
 F_i(T)\ =\ \int_0^{\frac{2}{5}T} dX\ X^{\epsilon_3}\ f_i(T,X)\ \Bigg|_{\epsilon_3=0} \ =\ \int_0^{\frac{2}{5}T} dX\ \Big( f_i(T,X) \ -\ \frac{(f_i)_{-2}}{X^2} \Big) \ +\ \int_0^{\frac{2}{5}T} dX\ X^{\epsilon_3}\ \frac{(f_i)_{-2}}{X^2}\ \Bigg|_{\epsilon_3=0} \ .
\end{equation}
It turns out that the coefficient of the $1/X^2$ pole, which in the above formula is indicated as $(f_i)_{-2}$, is $T$ independent.

\begin{figure}[h!]
  \centering
  \begin{subfigure}
   \centering
   \includegraphics[scale=.7]{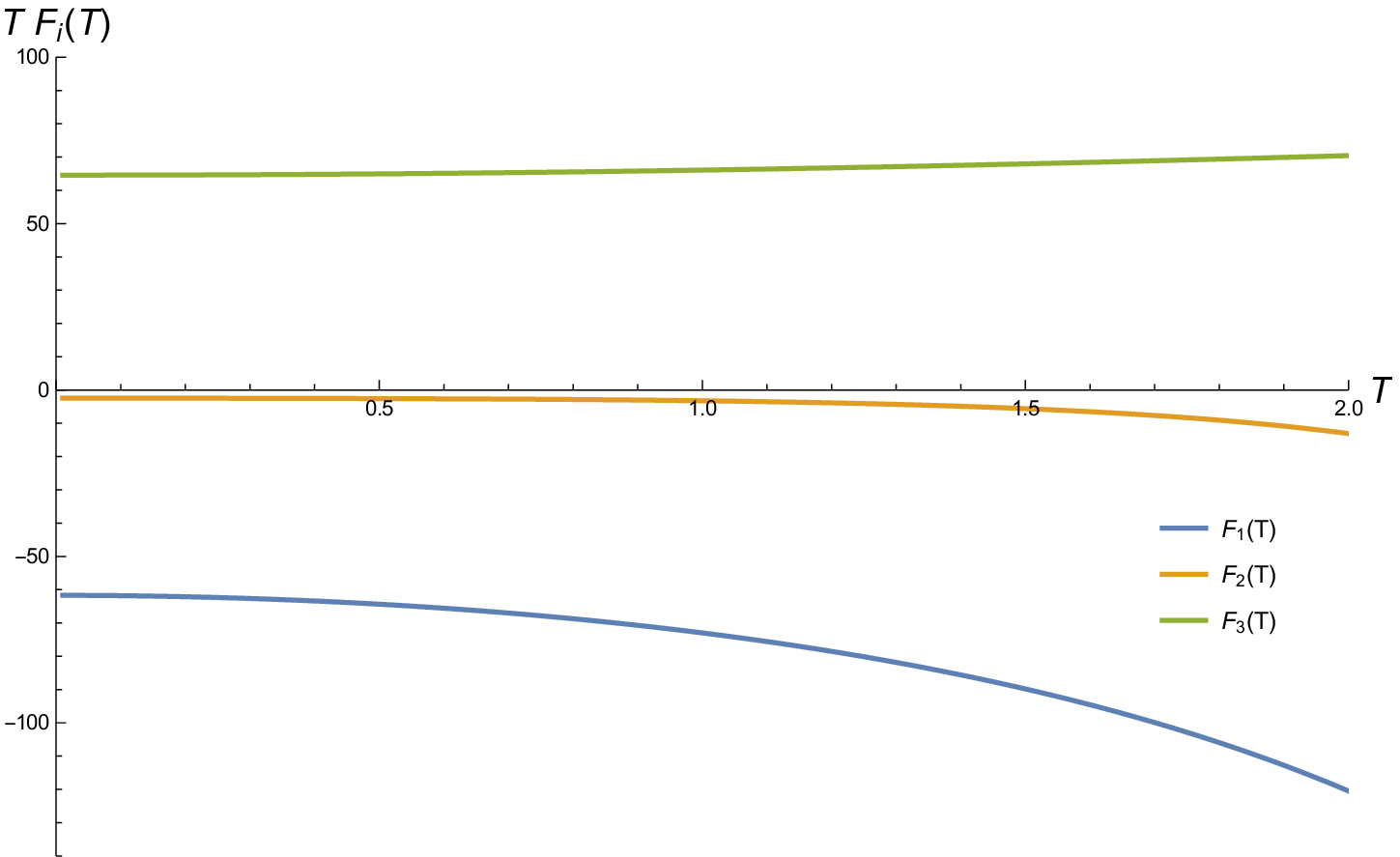} (a)
     \end{subfigure}%
  \begin{subfigure}
    \centering
    \includegraphics[scale=.7]{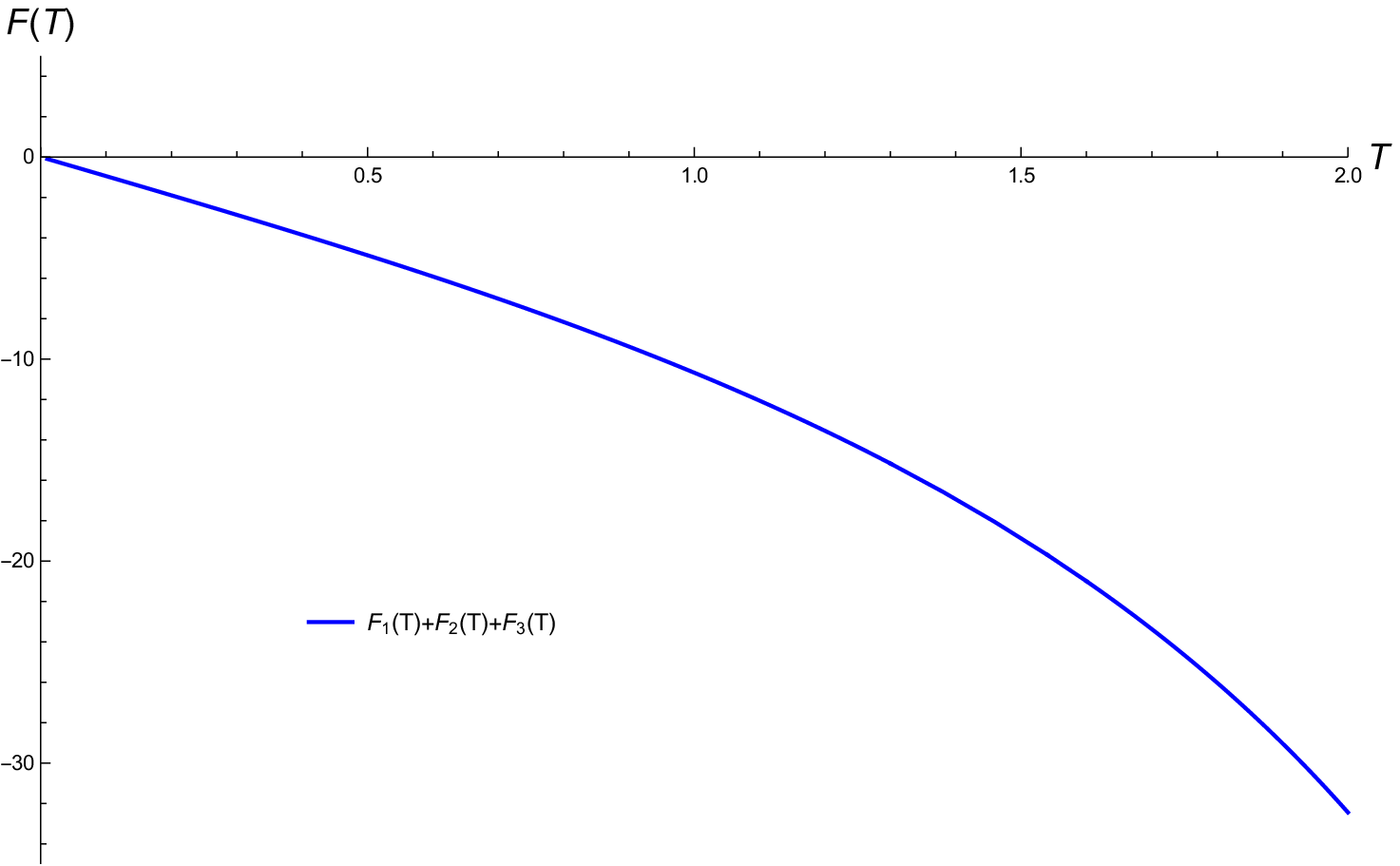}(b)
  \end{subfigure}
  \caption{ In (a) we show the three residues of the simple poles corresponding to the three diagrams. In (b) the sum of the three functions $F_i(t)$ is shown to be finite in $T=0$. \label{fig:div2} }
\end{figure}
This treatment leaves us with three numerical functions of $T$, which have to be integrated in the inteval $\left[0,\frac{5}{2}\right]$. Surprisingly each of these functions shows a nonvanishing $1/T$ pole, as shown in Figure \ref{fig:div1},
\begin{equation}
 F_i(T)\ =\ \frac{p_i}{T}\ +\ \tilde F_i(T) \ ,
\end{equation}
with $\tilde F_i$ finite in $T=0$. We explicitly find
\begin{equation}
 \begin{cases}
  p_1\ =\ -\ 62.10989(8) \\
  p_2\ =\ -\ 2.45519(8) \\
  p_3\ =\ 64.56508(9)
 \end{cases} \ .
\end{equation}
These poles are potentially problematic, and it is reassuring that the sum of them vanishes
\begin{equation}
 p_1\ +\ p_2\ +\ p_3 \ =\ 0 \ ,
\end{equation}
see also Figure \ref{fig:div2}. This is an important consistency check, because a $1/T$ pole would be an obstruction to the existence of the solution at the fifth order.
Numerically computing the integral over $T$ finally gives
\begin{equation}
 a_1 \ =\ 10.58226(7)\ .
\end{equation}
To conclude the total contribution to fifth order is given by
\begin{equation}
 B_{5,m}^{\tin{SFT}}\ =\ B_5^{\tin{BCFT}}\ +\ \lr{9 \log 2}\ B_{3,m}^{\tin{BCFT}}\ +\ 10.58226(7)\ B_{3,m}^{\tin{BCFT}}\ ,
\end{equation}
which corresponds to \eqref{comparing}
\begin{equation}
 b_5\ =\ 2.38996(7) \ .
\end{equation}

\section*{Acknowledgements}

We would like to thank Ted Erler,  Marialuisa Frau and Martin Schnabl for discussions. We especially thank Matej Kudrna for discussions and for sharing interesting Siegel gauge results in level truncation. PVL  warmly thanks Edoardo Lauria for useful discussions and comments. 
CM thanks Nathan Berkovits and the other organizers  of the VIII workshop {\it String Field Theory and Related Aspects}, where some of our preliminary results were reported. The research of CM is funded by a {\it Rita Levi Montalcini} grant from the Italian MIUR. 
This work is partially supported by the Compagnia di San Paolo contract {\it MAST: Modern Applications of String Theory} TO-Call3-2012-0088 and by the MIUR PRIN Contract 2015MP2CX4 {\it Non-perturbative Aspects Of Gauge Theories And Strings}.
\appendix

\section{Conventions and correlators}\label{app}

\paragraph{Witten's star product} $ $\\

\noindent
Witten star product is best understood in the sliver frame (with coordinate $z$), which is related to the UHP (with coordinate$w$) by the map
\begin{equation}
  z\ =\ \frac{2}{\pi}\ \arctan w \ .\label{sliverframe}
\end{equation}
In this frame the star product of wedge states with insertions is given by
\begin{equation}
 \begin{split}
  &\Bigg( U_{r}^* U_{r} \ \Phi_1( x_1)\ \ldots \ \Phi_n( x_n) \vk \Bigg) * \Bigg( U_{s}^*  U_{s}\ \Psi_1( y_1)\ \ldots \ \Psi_m( y_m) \vk \Bigg)  \\
  &=\ U_{r+s-1}^* U_{r+s-1}\ \Phi_1\lr{ x_1+\tfrac{s-1}{2}}\ \ldots \ \Phi_n\lr{ x_n+\tfrac{s-1}{2}} \ \Psi_1\lr{ y_1-\tfrac{r-1}{2}}\ \ldots \Psi_m\lr{ y_m-\tfrac{r-1}{2}} \vk \ ,
 \end{split}
 \label{starSchnabl}
\end{equation}
where  the coordinates are in the`` $2/\pi$'' sliver frame (\ref{sliverframe}), which implies some rescaling  wrt to (2.24) of  \cite{Schnabl:2005gv}, where the $2/\pi$ factor in (\ref{sliverframe}) was omitted.

\bigskip

\paragraph{Upper Half Plane}$ $\\

\noindent
In the BCFT of a free boson $X$ at the self-dual radius $R=1$, with Neumann boundary conditions, consider the bulk winding mode \begin{equation}
 \V(z,\z)\ =\ e^{i m \tilde X(z,\z)} \ ,
 \label{winding}
\end{equation}
where $\tilde{X}(z,\z) = X_L(z) - X_R(\z) $ is the T-dual field of $X(z,\z)$,
and the boundary marginal field
\begin{equation}
j(x)\ =\ i\sqrt2\partial X(x)=j(z){\Big| }_{z=x},
\end{equation} 
which is defined as a bulk chiral field placed at the boundary.
The chiral closed string field $X(z)$ has the following two-point functions ($\alpha'=1$)
\begin{equation}
 \left\{
  \begin{split}
   \corr{X(z) \ X(w)}_{\tin{ UHP }} \ &= \ - \dfrac{1}{2}\ \log (z-w) \\
   \corr{X(z) \ \p X(w)}_{\tin{ UHP }} \ &= \ \dfrac{1}{2}\ \dfrac{1}{z-w} \\
   \corr{ \p X(z) \ \p X(w)}_{\tin{ UHP }} \ &= \ -\ \dfrac{1}{2}\ \dfrac{1}{(z-w)^2}
  \end{split}
 \right. \ .
\end{equation}
The current $j(x)$ has the two point function
\begin{equation}
 \corr{j(x) \ j(y) }_{\tin{ UHP }} \ = \ \frac{1}{(x-y)^2} \ ,
\end{equation}
and it has the following OPE with the bulk winding mode $\mth{V}_m (z,\z) = e^{i m \tilde X(z,\z)}$ 
\begin{equation}
 \mth{V}_m (z,\z) \ j(x) \ \sim \ -\ \frac{m}{\sqrt{2}}\ \frac{1}{z-x}\ \mth{V}_m (z,\z)\ +\ \frac{m}{\sqrt{2}}\ \frac{1}{\z-x}\ \mth{V}_m (z,\z) \ ,
 \label{OPEV}
\end{equation}
We also have
\begin{equation}
 \corr{\tilde X(i,-i) \ \p X(x)}_{\tin{ UHP }} \ = \ -\ \frac{i}{1+x^2} \ ,
\end{equation}
and, using Wick theorem
\begin{equation}
 \corr{\mth{V}_m (i,-i) \ j(x) }_{\tin{ UHP }} \ = \  i m \sqrt{2} \ \frac{1}{1+x^2} \ \corr{\mth{V}_m (i,-i)}_{\tin{ UHP }} \ .
\end{equation}
The 1-point function is\footnote{This can be obtained from (A.1) of \cite{Kudrna:2012re} by mapping the UHP to the Disk, trading Dirichlet boundary conditions with Neumann and momentum with winding (T-duality) and setting $q=im$ there.}
\begin{equation}
 \corr{\mth{V}_m (i,-i)}_{\tin{ UHP }} \ = \ 2^{-m^2/2} \ .
\end{equation}
In addiction we have the following correlator for the auxiliary closed string field \cite{Kudrna:2012re}
\begin{equation}
 \corr{\mth{V}_{\tin{aux}}^{(1-h_m,1-h_m)}(i,-i)}_{\tin{ UHP }} \ =\ \frac{1}{4} \ 2^{m^2/2} \ .
 \label{normaux}
\end{equation}

\bigskip

%

\paragraph{Cylinder} $ $\\

\noindent
On the cylinder the chiral closed string field $X(z)$ has the following correlators \cite{Schnabl:2005gv}
\begin{equation}
 \left\{
  \begin{split}
   \corr{X(z) \ X(w)}_{C_N} \ &= \ -\ \dfrac{1}{2}\ \log \left[ \sin \left[ \dfrac{\pi}{N} (z-w) \right] \right] \\
   \corr{X(z) \ \p X(w)}_{C_N} \ &= \ \dfrac{\pi}{2N}\ \cot \left[ \dfrac{\pi}{N}  (z-w) \right] \\
   \corr{ \p X(z) \ \p X(w)}_{C_N} \ &= \ -\ \dfrac{\pi ^2}{N^2}\ \csc^2 \left[ \dfrac{\pi}{N}  (z-w) \right] \
  \end{split}
 \right. \ ,
 \label{greenfunctions}
\end{equation}
and then
\begin{equation}
 \corr{j(x) \ j(y) }_{C_N}\ = \ \dfrac{2\pi ^2}{N^2}\ \csc^2 \left[ \dfrac{\pi}{N}  (z-w) \right] \ .
 \label{2punti}
\end{equation}

%
%

\paragraph{Wick theorem}$ $\\
\label{sec:wick}

In the main text we deal with correlators of the following form
\begin{equation}
 \corr{\V(z,\z) \prod_{i=1}^N j_i(x_i) }\ =\ \corr{\V(z,\z)}\ \sum_{k=0}^N \frac{(im)^k}{k!}\ \corr{:(\tilde{X}(z,\z))^k: \prod_{i=1}^N j_i(x_i) } \ .
\end{equation}
This correlator significantly simplifies on a cylinder $C_n$, when the bulk operator (properly dressed with the ghosts and the auxiliary sector to acquire total weight zero, \eqref{VauxVmat}) is placed at the midpoint $i\infty$. 
In particular , thanks to the rotational invariance of $C_n$ we have 
\begin{equation}
 \corr{V_m(i\infty)\ c(x)}_{C_N} = \corr{V_m(i\infty)\ c(0)}_{C_N}\ =\ \frac{N}{\pi}\ \corr{V_m(i,-i)\ c(0)}_{UHP}\ =\ \frac{iN}{2\pi} \ .
 \label{ghostV}
\end{equation}
The term containing the maximal number of contractions with the closed string is given by
\begin{equation}
 \begin{split}
  \frac{(im)^N}{N!}\ \corr{c(0)\ V_m(i\infty)}\ &\corr{:(\tilde{X}(i\infty))^N: \prod_{i=1}^N j(x_i) } \ \\
  &=\ (im)^N\ \corr{c(0)\ V_m(i\infty)}\ \prod_{i=1}^N  \corr{\tilde{X}(i\infty) j(x_i) } \ \\
  &=\ (im)^N\ \corr{c(0)\ V_m(i\infty)}\ \prod_{i=1}^N  \corr{\tilde{X}(i\infty) j(0) } \ \\
  &=\  (im)^N\ \corr{c(0)\ V_m(i\infty)}\ \lr{  \corr{\tilde{X}(i\infty) j(0) }  }^N \ .
 \end{split}
\end{equation}
Let us list for convenience the explicit correlators which are used in the main text
\begin{equation}
 \begin{split}
  &\underline{N=0} \qquad \corr{c(0)\ V_m(i\infty)} \ , \\
  &\underline{N=1} \qquad in\ \corr{c(0)\ V_m(i\infty)} \ \corr{\tilde{X}(i\infty) j(0)} \ , \\
  &\underline{N=2} \qquad \corr{c(0)\ V_m(i\infty)} \ \Bigg\{ \corr{j(x_2)j(x_2)}\ -m^2\ \Big( \corr{\tilde{X}(i\infty) j(0)} \Big)^2 \Bigg\} \ , \\
  &\underline{N=3} \qquad \corr{c(0)\ V_m(i\infty)} \ \Bigg\{im\ \corr{\tilde{X}(i\infty) j(0)} \sum_{\substack{i,r=1\\ i<r}}^3 \corr{j(x_i)j(x_r)}\ -im^3\ \Big( \corr{\tilde{X}(i\infty) j(0)} \Big)^3 \Bigg\} \ , \\
  &\underline{N=4} \qquad \corr{c(0)\ V_m(i\infty)} \ \Bigg\{ \corr{j(x_1)j(x_2)j(x_3)j(x_4)} \  \\
  &\hspace{2cm}-\ m^2\ \Big( \corr{\tilde{X}(i\infty) j(0)} \Big)^2 \ \sum_{\substack{i,r=1\\ i<j}}^4 \corr{j(x_i)j(x_r)}\ +\ m^4\ \Big( \corr{\tilde{X}(i\infty) j(0)} \Big)^4 \Bigg\} \ , \\
  &\underline{N=5} \qquad \corr{c(0)\ V_m(i\infty)} \ \Bigg\{ im\ \Big( \corr{\tilde{X}(i\infty) j(0)} \Big)\ \corr{j(x_1)j(x_2)j(x_3)j(x_4)} \  \\
  &\hspace{2cm}-\ im^3\ \Big( \corr{\tilde{X}(i\infty) j(0)} \Big)^3 \ \sum_{\substack{i,r=1\\ i<r}}^5 \corr{j(x_i)j(x_r)}\ +\  im^5\ \Big( \corr{\tilde{X}(i\infty) j(0)} \Big)^5 \Bigg\} \ ,  \end{split}
 \label{Vamplitudes}
\end{equation}
where the only non-trivial  correlator involving $\tilde{X}$ is given by
\begin{equation}
 \corr{\tilde{X}(i\infty) j(0)}_{C_N} \ = \ \frac{\sqrt{2}\ \pi}{N} \ .
 \label{Vcilindro}
\end{equation}


\begin{thebibliography}{99}

\bibitem{Schnabl:2005gv}
  M.~Schnabl,
  ``Analytic solution for tachyon condensation in open string field theory,''
  Adv.\ Theor.\ Math.\ Phys.\  {\bf 10} (2006) no.4,  433
  doi:10.4310/ATMP.2006.v10.n4.a1
  [hep-th/0511286].

\bibitem{Schnabl:2007az}
  M.~Schnabl,
  ``Comments on marginal deformations in open string field theory,''
  Phys.\ Lett.\ B {\bf 654} (2007) 194
  doi:10.1016/j.physletb.2007.08.023
  [hep-th/0701248 [HEP-TH]].

\bibitem{KORZ}
  M.~Kiermaier, Y.~Okawa, L.~Rastelli and B.~Zwiebach,
  ``Analytic solutions for marginal deformations in open string field theory,''
  JHEP {\bf 0801} (2008) 028
  [hep-th/0701249 [HEP-TH]].

\bibitem{FKP}
  E.~Fuchs, M.~Kroyter and R.~Potting,
  ``Marginal deformations in string field theory,''
  JHEP {\bf 0709} (2007) 101
  doi:10.1088/1126-6708/2007/09/101
  [arXiv:0704.2222 [hep-th]].

\bibitem{KO}
  M.~Kiermaier and Y.~Okawa,
  ``Exact marginality in open string field theory: A General framework,''
  JHEP {\bf 0911} (2009) 041
  [arXiv:0707.4472 [hep-th]].

\bibitem{BMT}
  L.~Bonora, C.~Maccaferri and D.~D.~Tolla,
  ``Relevant Deformations in Open String Field Theory: a Simple Solution for Lumps,''
  JHEP {\bf 1111} (2011) 107
  [arXiv:1009.4158 [hep-th]].

\bibitem{Murata2}
  M.~Murata and M.~Schnabl,
  ``Multibrane Solutions in Open String Field Theory,''
  JHEP {\bf 1207} (2012) 063
  [arXiv:1112.0591 [hep-th]].

\bibitem{Erler:2012qn}
  T.~Erler and C.~Maccaferri,
  ``Connecting Solutions in Open String Field Theory with Singular Gauge Transformations,''
  JHEP {\bf 1204} (2012) 107
  doi:10.1007/JHEP04(2012)107
  [arXiv:1201.5119 [hep-th]].

\bibitem{Maccaferri:2014cpa}
  C.~Maccaferri,
  ``A simple solution for marginal deformations in open string field theory,''
  JHEP {\bf 1405} (2014) 004
  doi:10.1007/JHEP05(2014)004
  [arXiv:1402.3546 [hep-th]].

\bibitem{Erler:2014eqa}
  T.~Erler and C.~Maccaferri,
  ``String Field Theory Solution for Any Open String Background,''
  JHEP {\bf 1410} (2014) 029
  doi:10.1007/JHEP10(2014)029
  [arXiv:1406.3021 [hep-th]].


\bibitem{defects}
  T.~Kojita, C.~Maccaferri, T.~Masuda and M.~Schnabl,
  ``Topological defects in open string field theory,''
  arXiv:1612.01997 [hep-th].
  
\bibitem{Fuchs:2008cc}
  E.~Fuchs and M.~Kroyter,
  ``Analytical Solutions of Open String Field Theory,''
  Phys.\ Rept.\  {\bf 502} (2011) 89
  doi:10.1016/j.physrep.2011.01.003
  [arXiv:0807.4722 [hep-th]].
  
\bibitem{Schnabl:2010tb}
  M.~Schnabl,
  ``Algebraic solutions in Open String Field Theory - A Lightning Review,''
  Acta Polytechnica 50, no. 3 (2010) 102
  [arXiv:1004.4858 [hep-th]].
  
\bibitem{Okawa:2012ica}
  Y.~Okawa,
  ``Analytic methods in open string field theory,''
  Prog.\ Theor.\ Phys.\  {\bf 128} (2012) 1001.
  doi:10.1143/PTP.128.1001
  \bibitem{Sen:2000hx}
  A.~Sen and B.~Zwiebach,
  ``Large marginal deformations in string field theory,''
  JHEP {\bf 0010} (2000) 009
  doi:10.1088/1126-6708/2000/10/009
  [hep-th/0007153].

  
%
%

\bibitem{Karczmarek:2014wea}
  J.~L.~Karczmarek and M.~Longton,
  ``Renormalization schemes for SFT solutions,''
  JHEP {\bf 1504} (2015) 007
  doi:10.1007/JHEP04(2015)007
  [arXiv:1412.3466 [hep-th]].
  
\bibitem{Longton:2015maa}
  M.~Longton,
  ``Time-Symmetric Rolling Tachyon Profile,''
  JHEP {\bf 1509} (2015) 111
  doi:10.1007/JHEP09(2015)111
  [arXiv:1505.00802 [hep-th]].
  
\bibitem{Recknagel:1998ih}
  A.~Recknagel and V.~Schomerus,
  ``Boundary deformation theory and moduli spaces of D-branes,''
  Nucl.\ Phys.\ B {\bf 545} (1999) 233
  doi:10.1016/S0550-3213(99)00060-7
  [hep-th/9811237].

  
\bibitem{Maccaferri:2015cha}
  C.~Maccaferri and M.~Schnabl,
  ``Large BCFT moduli in open string field theory,''
  JHEP {\bf 1508} (2015) 149
  doi:10.1007/JHEP08(2015)149
  [arXiv:1506.03723 [hep-th]].

  
\bibitem{Kudrna:2016ack}
  M.~Kudrna and C.~Maccaferri,
  ``BCFT moduli space in level truncation,''
  JHEP {\bf 1604} (2016) 057
  doi:10.1007/JHEP04(2016)057
  [arXiv:1601.04046 [hep-th]].
  
\bibitem{KLM}
  M.~Kudrna, P.V.~Larocca and C.~Maccaferri,
  \textit{in progress}.

  
\bibitem{Kudrna:2012um}
  M.~Kudrna, T.~Masuda, Y.~Okawa, M.~Schnabl and K.~Yoshida,
  ``Gauge-invariant observables and marginal deformations in open string field theory,''
  JHEP {\bf 1301} (2013) 103
  doi:10.1007/JHEP01(2013)103
  [arXiv:1207.3335 [hep-th]].
  

  
\bibitem{Kudrna:2012re}
  M.~Kudrna, C.~Maccaferri and M.~Schnabl,
  ``Boundary State from Ellwood Invariants,''
  JHEP {\bf 1307} (2013) 033
  doi:10.1007/JHEP07(2013)033
  [arXiv:1207.4785 [hep-th]].
  
%
\bibitem{Ishibashi:1988kg}
  N.~Ishibashi,
  ``The Boundary and Crosscap States in Conformal Field Theories,''
  Mod.\ Phys.\ Lett.\ A {\bf 4} (1989) 251.
  doi:10.1142/S0217732389000320
  
\bibitem{Ellwood:2008jh}
  I.~Ellwood,
  ``The Closed string tadpole in open string field theory,''
  JHEP {\bf 0808} (2008) 063
  doi:10.1088/1126-6708/2008/08/063
  [arXiv:0804.1131 [hep-th]].

\bibitem{Berkovits:2003ny}
  N.~Berkovits and M.~Schnabl,
  ``Yang-Mills action from open superstring field theory,''
  JHEP {\bf 0309} (2003) 022
  doi:10.1088/1126-6708/2003/09/022
  [hep-th/0307019].
\bibitem{Schnabl:2002gg}
  M.~Schnabl,
  ``Wedge states in string field theory,''
  JHEP {\bf 0301} (2003) 004
  doi:10.1088/1126-6708/2003/01/004
  [hep-th/0201095].

  
  
  





  
  
  
  
  
  
  



  

\end{thebibliography}
\end{document}